\documentclass[11pt,letterpaper]{article}

\usepackage{latexsym,multirow}
\usepackage{amssymb,amsmath, bm, mathtools}
\usepackage{graphicx}
\usepackage[color,all,import,arrow]{xy}
\usepackage{enumerate}
\usepackage{setspace}
\usepackage{authblk}
\usepackage{xcolor}
\usepackage{bbm}
\usepackage{longtable}
\usepackage{array}
\usepackage{booktabs}

\usepackage{natbib}
\usepackage[pdftex, bookmarksopen=true, bookmarksnumbered=true,
pdfstartview=FitH, breaklinks=true, urlbordercolor={0 1 0}, citebordercolor={0 0 1}]{hyperref}
\usepackage{colortbl}
\usepackage{subfigure}

\usepackage{dcolumn}
\newcolumntype{.}{D{.}{.}{-1}}
\newcolumntype{d}[1]{D{.}{.}{#1}}
\usepackage{theorem}
\theoremstyle{plain}
\theoremheaderfont{\scshape}
\newtheorem{assumption}{Assumption}

\newtheorem{proposition}{Proposition}
\newtheorem{theorem}{Theorem}

\providecommand{\norm}[1]{\lVert#1\rVert}

\usepackage{rotating}

\usepackage[compact]{titlesec}

\allowdisplaybreaks

\newcommand{\pr}{\text{pr}}

\def\d{\textup{d}}

\newcommand*{\QEDB}{\null\nobreak\hfill\ensuremath{\square}}%

\def\T{\tiny\textsc{t}}

\def\sumn{\sum_{i=1}^n}

\def\d{\textnormal{d}}
\def\pr{\textup{pr}}
\def\c{\textup{cp}}
\def\Bern{\textup{Bernoulli}}

\newcommand{\indep}{\perp \!\!\! \perp}
\providecommand{\norm}[1]{\lVert#1\rVert}

\begin{document}
\doublespacing
\title{\bf \Large Estimating treatment effects with competing intercurrent events in randomized controlled trials}

\author[1]{Sizhu Lu}
\author[2]{Yanyao Yi}
\author[2]{Yongming Qu}
\author[2]{Huayu Karen Liu}
\author[3]{Ting Ye\thanks{Corresponding authors: Ting Ye and Peng Ding. Emails: \mbox{\texttt{tingye1@uw.edu}} and \mbox{\texttt{pengdingpku@berkeley.edu}}.}}
\author[1]{Peng Ding$^*$}

\affil[1]{Department of Statistics, University of California, Berkeley}
\affil[2]{Global Statistical Sciences, Eli Lilly and Company}
\affil[3]{Department of Biostatistics, University of Washington}

\date{\today}
\maketitle

\begin{abstract}
The analysis of randomized controlled trials is often complicated by intercurrent events (IEs) -- events that occur after treatment initiation and affect either the interpretation or existence of outcome measurements. Examples include treatment discontinuation or the use of additional medications. In two recent clinical trials for systemic lupus erythematosus with complications of IEs, we classify the IEs into two broad categories: effect-informative (e.g., treatment discontinuation due to adverse events or lack of efficacy) and effect-uninformative (e.g., treatment discontinuation due to external factors such as pandemics or relocation). To define a clinically meaningful estimand, we adopt tailored strategies for each category of IEs. For effect-informative IEs, which are often informative about a patient's outcome, we use the \emph{composite variable strategy} that assigns an outcome value indicative of treatment failure. For effect-uninformative IEs, we apply the \emph{hypothetical strategy}, assuming their timing is conditionally independent of the outcome given treatment and baseline covariates, and hypothesizing a scenario in which such events do not occur. A central yet previously overlooked challenge is the presence of competing IEs, where the first IE censors all subsequent ones. Despite its ubiquity in practice, this issue has not been explicitly recognized or addressed in previous data analyses due to the lack of rigorous statistical methodology. In this paper, we propose a principled framework to formulate the estimand, establish its nonparametric identification and semiparametric estimation theory, and introduce weighting, outcome regression, and doubly robust estimators. We apply our methods to analyze the two systemic lupus erythematosus trials, demonstrating the robustness and practical utility of the proposed framework.

\end{abstract}

\section{Intercurrent events in randomized controlled trials}
\label{sec::introduction}
Randomized controlled trials (RCTs) are considered the gold standard for evaluating treatment efficacy, primarily because randomization supports assumption-lean inference of the treatment effect. However, after treatment initiation, various events, referred to as intercurrent events (IEs), can arise, impacting the interpretation or availability of outcome measurements and posing significant challenges to the analysis of RCTs. Examples of IEs include treatment discontinuation due to adverse events or lack of efficacy, patient relocation, and the use of additional medications. Carefully accounting for IEs is essential to ensure the validity and reliability of the causal conclusions drawn from RCTs.

Recognizing the critical need for clearly defined estimands in the presence of IEs, the International Council for Harmonization (ICH) issued the E9(R1) Addendum \citep{international2019addendum}. This addendum introduces a structured estimand framework for clinical trials to obtain precisely defined treatment effects that align with the clinical questions of interest. It outlines strategies for addressing IEs during the formulation of the clinical question and emphasizes that careful specification of the treatment, population, and outcome variable often addresses many of the IEs raised in discussions between sponsors and regulators. Since its release, the ICH E9(R1) has been widely discussed, increasingly adopted in clinical drug development, and has sparked substantial interest in statistical research \citep{qu2021defining, kang2022incorporating, ionan2023clinical, han2023defining, olarte2025estimating}.

A widely accepted strategy is the \emph{treatment policy strategy} \citep{international2019addendum}, which includes all participants in their originally assigned groups and uses the observed outcome values, regardless of whether or not an IE occurs. This approach aligns with the intention-to-treat principle and reflects the effect of a treatment policy in real clinical settings. However, it cannot address IEs that are terminal events, such as death, because such events preclude the existence of the outcome variable.

The \emph{composite variable strategy} is an alternative approach that is well-suited for handling IEs that are informative of the patient's outcome such as the terminal events. This strategy incorporates the occurrence of IEs directly into the outcome definition. Specifically, it defines a composite outcome: if no IE occurs, the outcome of interest is used as observed; if an IE occurs, the endpoint is deterministically set to a pre-specified, clinically meaningful value, typically indicating treatment failure. This strategy is widely used in practice across various types of outcomes, including binary outcomes (e.g., the non-responder rule), ordinal or continuous outcomes (e.g., \citealp{rosenbaum2006comment}), and time-to-event outcomes (e.g., progression-free survival).

In addition to the treatment policy and composite variable strategies, \cite{international2019addendum} outlines three additional strategies for handling IEs. The \textit{hypothetical strategy} evaluates treatment effects under a hypothetical scenario in which the IE would not occur. The \emph{while-on-treatment} and \emph{principal stratification strategies} are also described but are less commonly used in current practice due to their reliance on strong assumptions and the potential to introduce bias in treatment comparisons. \cite{international2019addendum} recommends using different strategies based on the specific type of IEs involved. 

Several recent papers have begun to consider settings with multiple IE types and the use of multiple strategies. For example, \citet{lipkovich2020causal} and \citet{qu2021defining} discuss estimands that tailor strategies to different IEs without rigorously examining how to combine different strategies or rigorously addressing the potential pitfalls of doing so without careful consideration. More recently, \citet{olarte2025dealing} provides a careful conceptual discussion of estimands that combine the treatment policy and hypothetical strategies, using potential outcomes and causal diagrams to clarify alternative interpretations and the corresponding identification assumptions. Complementing this line of work, we study a different hybrid-strategy setting that combines the hypothetical strategy with the composite outcome strategy. Our setting also highlights a practically important feature in multi-IE settings that received limited explicit attention in the literature: multiple IEs can \emph{compete} with each other.

\subsection{Two phase-3 trials in systemic lupus erythematosus}
To rigorously examine the challenges posed by IEs in clinical trials, we analyze data from two recent randomized studies for systemic lupus erythematosus \citep{morand2023baricitinib, petri2023baricitinib}, 
in which participants were randomly assigned to either the baricitinib treatment group or the placebo control group. These twin trials were designed to provide substantiated evidence on the causal treatment effect of baricitinib versus placebo.

\begin{figure}[h]
    \centering
    \includegraphics[width=\linewidth]{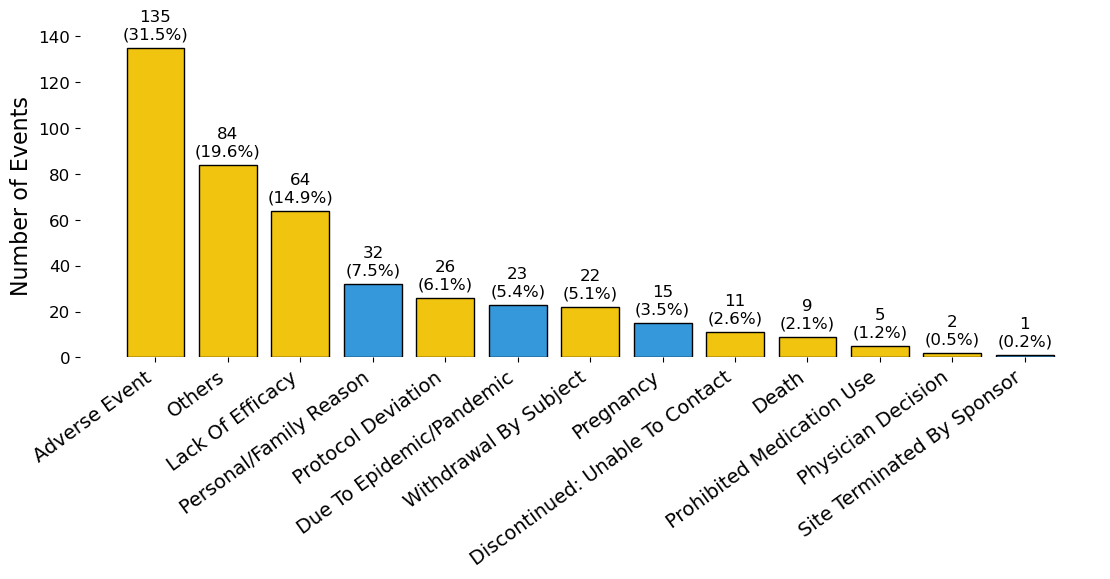}
    \caption{Bar plot showing the IE types and proportions. The yellow bars represent effect-informative IEs, and the blue bars represent effect-uninformative IEs.}
    \label{fig::pie_chart_ice}
\end{figure}

The primary outcome is a response index measured at 52 weeks after treatment initiation. Ideally, this outcome would be compared directly between the two treatment groups at week 52. However, 429 out of 1,535 patients (27.95\%) experienced IEs during the follow-up period, resulting in unobserved outcome data. Figure~\ref{fig::pie_chart_ice} shows the types and proportions of these IEs. During the 52 weeks, various IEs occurred: some patients discontinued treatment due to adverse events or lack of efficacy; others discontinued treatment due to study withdrawal for unspecified reasons or unable to contact; and some were excluded from the study due to protocol deviations. We revisit this example in greater detail in Section~\ref{sec::application}.

\subsection{Our proposal and contribution}
\label{sec::proposal_and_contribution}
We propose to classify IEs into two broad types: (1) effect-uninformative IEs, such as treatment discontinuation due to relocation or COVID-19 lockdown, which are assumed to be independent of treatment efficacy conditional on the observed covariates; and (2) effect-informative IEs, such as treatment discontinuation due to adverse events or lack of efficacy, use of rescue medication, and terminating events such as death, which are often informative about a patient’s treatment effect. 

In our application studies, the two types of IEs were classified based on a manual review of the detailed comments collected at the clinical sites. Figure~\ref{fig::pie_chart_ice} summarizes the main categories and illustrates our classification scheme: yellow bars represent effect-informative IEs, and blue bars represent effect-uninformative IEs. For instance, some patients withdrew from the study due to external factors such as relocation or the COVID-19 pandemic, which were classified as effect-uninformative. Others withdrew due to concerns about potential side effects or lack of efficacy, which were considered effect-informative. When the reason for withdrawal was unclear in the documentation, we defaulted to classifying the event as effect-informative and, under our composite outcome construction, assigned the corresponding failure value. This choice is intended to avoid overestimating arm-specific mean outcomes, while it does not necessarily make the estimated treatment effect conservative, since the net direction depends on how often such cases occur in each arm. Both types of IEs may occur during the follow-up period. A complete list of the IE categories observed, together with their assigned classification and rationale, is provided in Supplementary Section~\ref{sec::supp_ie_classification}.

\begin{figure}[t]
    \centering
    \includegraphics[width=.75\linewidth]{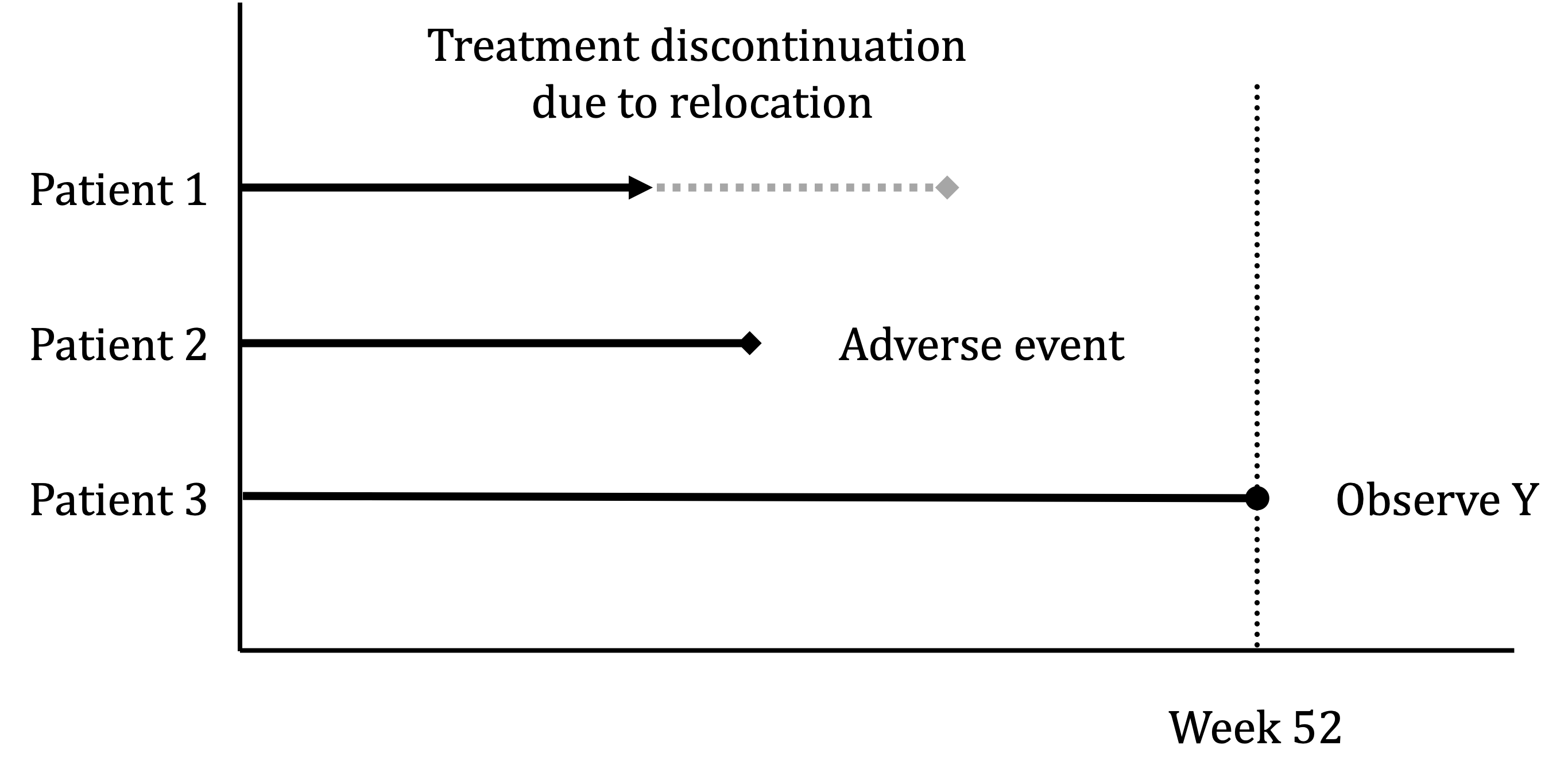}
    \caption{Illustration of three representative scenarios of intercurrent events in the immunology trial. The primary outcome is measured at week 52, and the solid lines indicate the observed follow-up period for each patient. Patient 1 experienced an effect-uninformative IE (represented by an arrow), Patient 2 experienced an effect-informative IE (represented by a diamond), and Patient 3 had no IE before week 52. This figure highlights the issue of competing IEs: for instance, had Patient 1 not discontinued early and remained on treatment (illustrated by the gray dashed line), they might have experienced an effect-informative IE later in the follow-up period.
    }
    \label{fig::example_illustration}
\end{figure}

Figure~\ref{fig::example_illustration} depicts the timeline and three representative scenarios from our motivating example. Before week 52, patient 1 withdrew from the study due to relocation, which is plausibly unrelated to the biological effect of treatment. Patient 2 discontinued due to an adverse event, reflecting an effect-informative IE. Patient 3 completed the study through week 52, at which point the primary outcome was measured.

Scenarios in Figure~\ref{fig::pie_chart_ice} also highlight why a single, uniform estimand strategy can be misaligned with the clinical decision problem. Both types of IEs can result in an unmeasured primary outcome at the pre-specified landmark time: once an IE occurs, the outcome may be unobserved, and in some effect-informative cases such as death, it may not even be well-defined. If we apply the hypothetical strategy to all IEs, we implicitly ask an unrealistic question: ``Does the drug work in a world where no one ever experienced clinically meaningful failures such as adverse events or lack of efficacy?'' In reality, what decision-makers care about instead is: ``Should we approve this drug, accounting for both possible benefit and clinically meaningful failures?'' Conversely, if we apply a composite outcome strategy to all IEs, we would incorrectly penalize external disruptions such as relocation, and the resulting estimate would carry unnecessarily little information. Moreover, answering the unrealistic questions using a hypothetical strategy to all IEs typically requires untestable, strong, and thus hard-to-defend assumptions such as conditional independence between the time to an adverse event and the potential outcomes of interest given baseline covariates. Related concerns arise for principal stratification estimands, which target effects in latent subgroups and generally rely on untestable assumptions for identification. By contrast, our combined strategy is designed to align the estimand with regulatory decision questions while relying on more transparent assumptions. We formalize these assumptions and the corresponding identification results in Section~\ref{sec::identification}.

Motivated by this mismatch and to ensure the estimand is clinically interpretable and relevant, we apply different strategies tailored to the nature of the IE. 
For IEs that are plausibly unrelated to the biological effect of the treatment (e.g., treatment discontinuation due to relocation or administrative withdrawal), we apply the \emph{hypothetical strategy}, which imagines a scenario where the IE did not occur. This enables estimation of the treatment effect as if the patient had remained on treatment and in the study as planned, under the assumption that the IE time is conditionally independent of the outcome given treatment and baseline covariates. This assumption is reasonable in settings where the IE arises from external or administrative factors that are not influenced by post-randomization conditions. For example, if a patient relocates for personal reasons unrelated to their health status or treatment response, then, conditional on baseline characteristics and assigned treatment, the occurrence of this IE can be viewed as independent of the patient's outcome.
In contrast, for effect-informative IEs (e.g., treatment discontinuation due to adverse events or lack of efficacy), we apply the \emph{composite variable strategy}, treating the IE itself as an indication of treatment failure, as such events signify that the patient is unable to continue treatment. A formal definition and discussion of this strategy are provided in Section~\ref{sec::identification}. Combining the two strategies enhances the interpretability of the estimand and ensures it better reflects clinically relevant questions, as no single strategy can adequately address both types of IEs. Specifically, it would be inappropriate to classify a patient as a non-responder if they discontinue treatment due to relocation, and it is of limited clinical relevance to consider a hypothetical scenario in which adverse events do not occur. This approach is consistent with regulatory guidance, which recommends applying different strategies depending on the type of IE involved  \citep{international2019addendum,kang2022incorporating,kahan2024estimands}.

However, a central yet previously overlooked challenge is the presence of \emph{competing IEs}, where the first IE censors the subsequent
ones. For example, if a patient discontinues treatment due to relocation, as illustrated by patient 1 in Figure~\ref{fig::example_illustration}, we only observe that no effect-informative IE occurred before the discontinuation. However, it remains unknown whether an adverse event would have occurred before the final outcome measurement in the hypothetical scenario in which the patient had not relocated and had continued treatment as planned. Although such competing IEs are ubiquitous in practice, this issue has not, to our knowledge, been explicitly recognized or addressed in prior methodological work. A related recent paper, \citet{olarte2025dealing}, also considers settings with two types of IEs whose causal ordering matters, and distinguishes different cases using directed acyclic graphs and single-world intervention graphs. Our focus is different. Rather than analyzing these through a graphical causal framework, we adopt a survival analysis and competing event perspective and explicitly model the event times of both IEs, which allows us to address settings in which one IE censors the observation of another.

Therefore, the central thesis of this paper is to address the challenge of competing IEs. We begin by developing a principled framework that clearly defines the estimand and establishes its nonparametric identification. Specifically, we derive two identification formulas, each relying on a different set of nuisance parameters. Building on these results, we propose two basic estimators corresponding to the two identification strategies. We then introduce an augmented estimator that combines the two, achieving double robustness. To further enhance both robustness and efficiency, we derive the efficient influence function (EIF) and construct an EIF-based estimator that attains the semiparametric efficiency bound under appropriate conditions. We apply our methods to the two systemic lupus erythematosus trials, both partially impacted by the COVID-19 pandemic,
demonstrating the robustness and practical utility of the proposed framework.

\subsection{Organization and notation}

The remainder of the paper is organized as follows. In Section~\ref{sec::setup}, we present a motivating example, introduce the basic setup of our research question, define the causal parameter of interest, and highlight the identification challenge. In Section~\ref{sec::identification}, we state the key identification assumptions, establish the nonparametric identification of the causal parameter, and construct three estimators. In Section~\ref{sec::eif}, we derive the EIF, propose an estimator that is both doubly robust and asymptotically efficient, and examine its asymptotic properties. In Section~\ref{sec::application}, we apply the methods to re-analyze the two systemic lupus erythematosus trials. Finally, in Section~\ref{sec::discussion}, we conclude with a discussion of two directions for future research. The supplementary material includes a simulation study and all proofs. We also provide publicly available R code implementing all four proposed estimators.

We use the following notation. Let $\norm{r}_2 = \{\int r(o)^2 \d P(o)\}^{1/2}$ denote the $L_2(P)$ norm where $P(\cdot)$ denotes the distribution of the observed data $O=o$. For the survival functions, let $\norm{r}_2 = \{\iint r(t,o)^2\d P(o)\d t\}^{1/2}$ denote the $L_2(P)$ norm. We write $b_n=O_P(a_n)$ if $b_n/a_n$ is bounded in probability and $b_n=o_P(a_n)$ if $b_n/a_n$ converges to $0$ in probability.

\section{Setup and estimand}
\label{sec::setup}

\subsection{Setup}
Let $A$ denote the binary treatment indicator, where $A = 1$ corresponds to assignment to the treatment group and $A = 0$ to the control group. The primary outcome, denoted by $Y$, is measured at a pre-specified time point $k$. In our motivating example, $Y$ is the response index and $k$ is 52 weeks. We adopt the potential outcomes framework, and write $Y(a)$ for the potential outcome under treatment assignment $A=a$, for $a\in\{0,1\}$.

As discussed in Section~\ref{sec::introduction}, two types of IEs may occur after treatment initiation: effect-uninformative IEs and effect-informative IEs. Let $C$ denote the time to an effect-uninformative IE and $T$ denote the time to an effect-informative IE. Because both are post-treatment variables, we write $C(a)$ and $T(a)$ for their potential values under treatment assignment $A=a$, where $a\in\{0,1\}$. Since our interest is restricted to events occurring before the outcome assessment time $k$, we set $C(a)=\infty$ if $C(a)>k$ and $T(a)=\infty$ if $T(a)>k$. The observed event times are therefore $C = C(A) = A C(1) + (1-A) C(0)$ and $T = T(A) = A T(1) + (1-A) T(0)$.
In the observed data, we do not separately observe both post-treatment event times for the same individual. Rather, we only observe the earliest of the two event times before $k$, $C(A)\wedge T(A)\wedge k$. Let $X$ denote the vector of collected baseline covariates.

Our estimand is motivated by the hypothetical strategy for effect-uninformative IEs, under which we ask what would have happened had such an IE not occurred. This indicates that $C$ lies on the causal pathway from treatment to the outcome. We therefore introduce the nested potential outcome $Y(a,c)$, which denotes the outcome under treatment assignment $A=a$ if the effect-uninformative IE occurs at time $c$, with $c=\infty$ corresponding to no effect-uninformative IE before time $k$. For example, suppose a treated patient experiences an effect-uninformative IE at week 20 and consequently discontinues treatment. We will not observe their outcome of interest. However, even if we could send the patient back to the clinic at week 52 and get their primary outcome measured, that observed quantity would correspond to $Y(1,c=20)$ rather than to the hypothetical outcome of interest, $Y(1,c=\infty)$, which represents the outcome that would have been observed had the effect-uninformative IE not occurred before time $k$.

\subsection{Causal estimand and challenges in identification}
\label{subsec::causal_parameter}


We address the effect-informative IEs using the composite outcome strategy and address the effect-uninformative IEs using the hypothetical strategy. To combine them, we formally define a combined potential outcome as
\begin{eqnarray*}
    Y^{*}(a) &=& Y(a,c=\infty)1\{T(a)>k\} \ =\ 
    \begin{cases}
    Y(a, c=\infty) & \text{if } T(a) > k, \\
    0              & \text{if } T(a) \le k .
    \end{cases}
\end{eqnarray*}

This construction defines the outcome as zero (i.e., failure) if an effect-informative IE occurs before the outcome is measured. We then define our causal estimand as the mean contrast:
\begin{equation}
\label{eqn::tau_def}
    \tau = E\{Y^\c(1) - Y^\c(0)\} = E[Y(1,c=\infty)1\{T(1)>k\}] - E[Y(0,c=\infty)1\{T(0)>k\}].
\end{equation}
This estimand addresses the effect-informative IEs using the \emph{composite variable strategy}. It is widely used for binary outcomes to represent treatment success or failure, where effect-informative IEs such as adverse events or lack of efficacy are treated as failures \citep{international2019addendum}. In our motivating example, $Y(a,c=\infty)$ indicates whether a patient would be a responder at week 52 under treatment assignment $A = a$, had no IE occurred. For patients who experience an effect-informative IE before week 52, $Y(a,c=\infty)$ is unobserved. Under the composite variable strategy, these patients are defined as non-responders.

The composite variable approach extends beyond binary outcomes, but its applicability depends on whether treatment failure can be represented by a clinically meaningful reference value. For ordinal or categorical outcomes, this is often feasible when there is a clearly defined worst category or failure category. For continuous outcomes, the approach is most compelling when a prespecified value $v$ has a clear clinical interpretation as no benefit or treatment failure. Specifically, for outcomes such as chronic pain, physical functioning, and cognitive performance that are typically measured on an ordinal or continuous scale, if a predefined value $v$ may be used to indicate treatment failure, then a composite outcome under treatment assignment $A = a$ can be constructed as $Y(a,c=\infty) 1\{T(a)>k\} + v 1\{T(a)\leq k\}$, where the outcome retains its actual value if no effect-informative IE occurs before time $k$, and takes the failure value $v$ otherwise. The corresponding causal estimand can then be defined as in~\eqref{eqn::tau_def}, with $Y(a,c=\infty)$ replaced by $Y(a,c=\infty) - v$ for $a = 0,1$. For example, in settings where $Y$ is percent weight loss from baseline, $v=0$ could correspond to no weight loss, or where $Y$ is improvement from baseline on a symptom score, $v=0$ could correspond to no improvement. 

At the same time, we acknowledge that for some continuous outcomes the choice of a single failure value may be less obvious and may not be universally acceptable. \cite{rosenbaum2006comment} discussed a similar causal parameter in settings where the IE is death and the outcome of interest is a measure of quality of life assessed after a fixed period, where there may be no unique value that fully captures the clinical meaning of an effect-informative IE. In such settings, the composite variable strategy remains a possible estimand choice, but its interpretability depends critically on whether the chosen reference value is clinically justified in advance. When no such value is well defined, the applicability of this strategy is more limited. 

Identifying $\tau$ in the presence of both types of IEs presents a key methodological challenge. For clarity, in the remainder of this subsection, we use treatment discontinuation due to relocation and adverse events to represent effect-uninformative and effect-informative IEs, respectively. The causal estimand $\tau$ defined in~\eqref{eqn::tau_def} is based on the composite potential outcome $Y^\c(a)$, which takes the value $0$ if an adverse event occurs before the outcome measurement, i.e., when $T(a) < k$. If adverse event were the only type of IE, and treatment is randomized, $\tau$ could be identified from observed data using the difference in means of the composite outcomes: $\tau = E\{Y1(T>k)\mid A=1\} - E\{Y1(T>k)\mid A=0\}$. However, this formula is infeasible in the presence of effect-uninformative IEs, which may occur before effect-informative IEs and thus censor both $T(a)$ and $Y(a,c=\infty)$. As illustrated in Figure~\ref{fig::example_illustration}, patient 1 discontinued due to relocation, which censored both the occurrence of a potential adverse event and the outcome, resulting in an unobserved composite outcome $Y {1}(T > k)$. Therefore, for patients who experience effect-uninformative IEs, the problem cannot be treated as standard censoring, and one cannot na\"ively impute the outcome $Y$ under a hypothetical scenario where the effect-uninformative IE did not occur. This is because such patients may still have experienced an effect-informative IE had they not discontinued due to relocation. In other words, a valid identification strategy must recover the expected value of the composite outcome $Y {1}(T > k)$, rather than the outcome $Y$ alone.

Table~\ref{tab::summary_ice_outcome} summarizes the observed IE types and corresponding composite outcomes, highlighting that competing IEs complicate the identification of $\tau$. When no IE occurs, the outcome $Y$ is observed and equals the composite outcome since $T > k$. When an adverse event is observed, the composite outcome is, by definition, 0. However, when a patient discontinues treatment due to relocation, both the outcome $Y$ and the time to the adverse event $T$ are unobserved, and so is the composite outcome. We address this challenge and present formal identification results in the following section.

\begin{table}[h]
\caption{Summary of the observed IE types and outcome. $C$ is the time to treatment discontinuation due to relocation, $T$ is the time to adverse event, $Y$ is the outcome of interest, and $k$ is the pre-specified time point when the measurement of $Y$ is taken. A question mark ``?'' indicates that the corresponding value is unobserved.}
\doublespacing
\centering
\begin{footnotesize}
\begin{tabular}{lcccc}
\\
\hline
\hline
observed IE type &  $(T,C,k)$-relationship & $Y$& $1(T>k)$& composite outcome $Y^{\c}$ \\ 
\hline 
no any type of IE & $ C \wedge T>k$& $Y$ & 1&$Y$\\
adverse event & $ C \wedge k > T$ &  ?&0& 0 \\
treatment discontinuation due to relocation & $T \wedge k > C$ & ? &?&?\\
\hline 
\end{tabular}
\end{footnotesize}
\label{tab::summary_ice_outcome}
\end{table}

\section{Nonparametric identification and basic estimators}
\label{sec::identification}

\subsection{Assumptions}
We assume that the joint distribution of $\{X_i, A_i, T_i(1), T_i(0), C_i(1), C_i(0), Y_i(1,c=\infty), Y_i(0,c=\infty)\}$ for patient $i$ is independently and identically distributed from a superpopulation. For notational simplicity, we omit the subscript $i$ when there is no confusion.

We begin by stating the following assumption on treatment assignment.
\begin{assumption}[Treatment assignment] 
\label{assump::randomization}
We assume the following conditional independence and overlap conditions:
\begin{itemize}
    \item[(a)] $A \indep \{Y(a,c=\infty), T(a), C(a)\} \mid X $ for $a = 0,1$.
    \item[(b)] For some constant $\eta\in(0,0.5)$, $\eta < e(X) < 1-\eta$ with probability 1, where $e(X)=\pr(A=1\mid X)$ denotes the propensity score \citep{rosenbaum1983central}.
\end{itemize}    
\end{assumption}

Assumption~\ref{assump::randomization} holds in RCTs by design, allowing our results to apply directly to settings such as our motivating example. However, our formulation is more general and can also accommodate cases where treatment assignment is not completely randomized, such as in stratified randomized experiments and observational studies. 

Under Assumption~\ref{assump::randomization}, the causal estimand $\tau$ can be expressed as:
\begin{eqnarray*}
    \tau &=& E[E\{Y1(T>k)\mid A=1, X\} - E\{Y1(T>k)\mid A=0, X\}] \\
    &=& E\left\{\frac{AY1(T>k)}{e(X)} - \frac{(1-A)Y1(T>k)}{1-e(X)}\right\},
\end{eqnarray*}
which correspond to the standard outcome regression and inverse probability weighting identification formulas, respectively, when treating the combined potential outcome $Y^\c(a) = Y(a,c=\infty){1}\{T(a) > k\}$ as the ``new" outcome of interest. However, as discussed in Section~\ref{subsec::causal_parameter}, these expressions no longer serve as feasible identification formulas in the presence of competing effect-uninformative IEs. This is because $Y {1}(T > k)$ is not fully observed if an effect-uninformative IE occurs.

To address this challenge, we introduce additional assumptions on effect-uninformative IEs. Specifically, we assume that the potential time of an effect-uninformative IE under treatment $a$ is conditionally independent of the potential outcomes and the potential time to effect-informative IEs, given baseline covariates. This assumption is analogous to the censoring-at-random assumption in survival analysis when treating effect-uninformative IEs as censoring events \citep{robins1994estimation,tsiatis2006semiparametric}. Additionally, we also assume a standard positivity condition ensuring that the probability of no occurrence of effect-uninformative IE by time $k$, given observed covariates, is strictly positive. These conditions are summarized in the following assumption.
\begin{assumption}[Effect-uninformative IE]
\label{assump::car}
We assume the following conditional independence and positivity conditions for the time of effect-uninformative IE:
\begin{itemize}
    \item[(a)] $C(a) \indep \{Y(a,c=\infty), T(a)\} \mid X$ for $a = 0,1$.
    \item[(b)] For some constant $\eta_C>0$, $\pr\{C(a)>k\mid X\} > \eta_C$ with probability 1.
\end{itemize} 
\end{assumption}

Assumption~\ref{assump::car}(a) states that the time to an effect-uninformative IE, such as treatment discontinuation due to relocation, is independent of both the potential outcome (i.e., whether a patient would be a responder at week 52) and the time to an effect-informative IE (e.g., an adverse event), conditional on observed baseline covariates. We also emphasize that this is not a blanket assumption for all IEs. Rather, it is intended only for events that, after clinical review, are judged to be plausibly unrelated to the patient's underlying biological response to treatment. This assumption is most credible for IEs driven by external or administrative factors, such as relocation for personal or work reasons, pandemic-related disruptions, travel restrictions, or site closure for administrative reasons. In such settings, once we condition on treatment assignment and relevant baseline characteristics, the occurrence time of the event may reasonably be viewed as carrying little additional information about the patient's latent outcome or latent time to an effect-informative IE. By contrast, the assumption is less credible for discontinuations related to adverse events, worsening symptoms, perceived lack of efficacy, or other evolving post-randomization clinical factors. Such events are therefore classified as effect-informative and incorporated into the composite outcome rather than handled under the hypothetical strategy. Thus, Assumption~\ref{assump::car}(a), while strong and untestable from observed data, provides a clear and clinically motivated criterion for classifying a subset of IEs as effect-uninformative.

Assumption~\ref{assump::car}(b) requires that every patient has a strictly positive probability of not experiencing an effect-uninformative IE before time $k$, given their covariates. This standard positivity condition rules out deterministic effect-uninformative IE happening in any subpopulation and is reasonable in our application.

If richer longitudinal covariates are available, Assumption~\ref{assump::car}(a) may be relaxed to a history conditional independence condition, in which $C(a)$ is assumed independent of $\{Y(a,c=\infty),T(a)\}$ conditional on the history of all time-varying covariates. In applications, it is also useful to assess robustness under alternative clinically meaningful classifications of ambiguous IE categories or through models that parameterize departures from conditional independence. We revisit the former direction in Section~\ref{sec::robustness_checks} and leave detailed methodologies of the latter extension for future work.

\subsection{Two identification formulas}
Using the composite variable strategy to handle effect-informative IEs, recall that we define the causal estimand as in~\eqref{eqn::tau_def}. Under Assumptions~\ref{assump::randomization} and~\ref{assump::car}, we show that the estimand $\tau$ is nonparametrically identifiable and present two identification formulas in the following theorem.

\begin{theorem}[Nonparametric identification of $\tau$]
\label{thm::comp_out_identification}
Under Assumptions \ref{assump::randomization} and~\ref{assump::car}, $\tau$ is nonparametrically identified by two distinct formulas. First,
\begin{eqnarray}
    \tau &=& E\left\{ \mu_1(X) S_1(k\mid X) - \mu_0(X) S_0(k\mid X) \right\}, \label{eqn::id_out}
\end{eqnarray}
where for $a=0,1$, $\mu_a(X)=E(Y\mid T\wedge C>k, X, A=a)$ is the conditional mean of observed outcome among those with no IE and $A=a$, and $S_a(t\mid X)= \pr(T>t\mid X, A=a)$ is the conditional survival function of effect-informative IE in treatment arm $a$. Second,
\begin{eqnarray}
    \tau &=& E\left[ \frac{AY1(T\wedge C>k)}{e(X)G_1(k\mid X)} - \frac{(1-A)Y1(T\wedge C>k)}{\left\{1-e(X)\right\}G_0(k\mid X)} \right] \label{eqn::id_ipw},
\end{eqnarray}
where $e(X)=\pr(A=1\mid X)$ is the propensity score and for $a=0,1$, $G_a(t\mid X) = \pr(C>t\mid X, A=a)$ is the conditional survival function of effect-uninformative IE in treatment arm $a$. 
\end{theorem}

The first identification formula~\eqref{eqn::id_out} expresses $\tau$ using a standard outcome regression approach. It decomposes the conditional expectation of the composite outcome given covariates and treatment as $E\{Y 1(T>k) \mid X, A \}= E(Y  \mid T>k, X, A ) \pr(T>k\mid X, A) = E(Y  \mid T\wedge C>k, X, A ) \pr(T>k\mid X, A) $. This formulation models the conditional mean outcome $Y$ among individuals who do not have any IE by time $k$, and weights it by the probability of not experiencing an effect-informative IE by that time. The second identification formula~\eqref{eqn::id_ipw} provides an alternative identification strategy based on inverse probability weighting. It identifies $\tau$ by reweighting observed outcomes among individuals who remain free of IEs up to time $k$, separately within each treatment arm. The weights adjust for the different probability of observing the composite outcome and consist of two components: $e(X)$, which accounts for differences in treatment assignment probabilities, and $G_a(k\mid X)$, which adjusts for the probability of remaining free of effect-uninformative IEs up to time $k$ within each treatment group. Effectively, individuals with a lower probability of being observed are up-weighted, while those with a higher probability are down-weighted, ensuring an unbiased estimate of the treatment effect.

Both equations~\eqref{eqn::id_out} and~\eqref{eqn::id_ipw} are valid identification formulas, as the nuisance parameters involved are either functions of observed data or identifiable from observed data. Specifically, the propensity score  $e(X)$ and the conditional outcome model $\mu_a(X)$ are functions of observed data. Although the survival functions for both types of IEs, $S_a(t\mid X)$ for effect-informative IEs and $G_a(t\mid X)$ for effect-uninformative IEs, are not direct functions of the observed data, they are identifiable under standard results in survival analysis with censoring at random \citep{robins1992recovery, robins2000correcting,ebrahimi2003identifiability}. For example, $S_1(t\mid X)=\pr(T>t\mid X, A=1)$ for $t\leq k$ can be estimated using data from the treatment group by viewing $T\wedge C\wedge k$ as the observed event time and using $1(C\wedge k\geq T)$ as the event indicator for the effect-informative IE. Estimation can then proceed using parametric or semiparametric methods, such as the Cox proportional hazards model \citep{cox1972regression}, or more flexible approaches \citep{wolock2024framework}. A similar approach can be used to estimate $G_1(t\mid X)$ by switching the roles of $T$ and $C$. The corresponding functions $S_0(t\mid X)$ and $G_0(t\mid X)$ can be estimated in the same way using data from the control group. 

\subsection{Two basic estimators}
The identification formulas~\eqref{eqn::id_out} and~\eqref{eqn::id_ipw} in Theorem \ref{thm::comp_out_identification} motivate two basic estimators, each relying on estimates of the corresponding nuisance functions. Let $\hat{S}_a(t\mid X)$ and $\hat{G}_a(t\mid X)$ denote the fitted survival models for effect-informative and effect-uninformative IEs, respectively, for $t\leq k$ and $a=0,1$. Let $\hat{\mu}_a(X)$ be the fitted outcome regression model for each treatment arm, and $\hat{e}(X)$ be the fitted propensity score model. By substituting these fitted values into the identification formulas and taking empirical analogs, we obtain the following two estimators:
\begin{eqnarray*}
    \hat{\tau}^{\textup{out}} &=& n^{-1}\sumn \hat{\mu}_1(X_i) \hat{S}_1(k\mid X_i) - n^{-1}\sumn \hat{\mu}_0(X_i) \hat{S}_0(k\mid X_i), \\
    \hat{\tau}^{\textup{ipw}} &=& n^{-1}\sumn \frac{A_i Y_i 1(T_i \wedge C_i > k)}{\hat{e}(X_i)\hat{G}_1(k\mid X_i)} - n^{-1}\sumn \frac{(1-A_i) Y_i 1(T_i \wedge C_i > k)}{\left\{1-\hat{e}(X_i)\right\}\hat{G}_0(k\mid X_i)}.
\end{eqnarray*}
The outcome regression estimator $\hat{\tau}^{\textup{out}}$ is consistent if both $\hat \mu_a(X)$ and $\hat{S}_a(k \mid X)$ are consistently estimated. The inverse probability weighting estimator $\hat{\tau}^{\textup{ipw}}$ is consistent if both $\hat{e}(X)$ and $\hat{G}_a(k \mid X)$ are consistently estimated.

\subsection{An augmented estimator}
\label{subsec::ps_aug_est}

Motivated by combining the two identification formulas~\eqref{eqn::id_out} and~\eqref{eqn::id_ipw}, we introduce another estimator that augments the weighting estimator $\hat\tau^{\textup{ipw}}$ using the estimated outcome models,
\begin{eqnarray*}
    \hat\tau^{\textup{aug}} &=& \hat\tau^{\textup{ipw}} - n^{-1} \sumn \left\{\frac{A_i-\hat e(X_i)}{\hat e(X_i)}\hat\mu_1(X_i)\hat S_1(k\mid X_i) + \frac{A_i -\hat e(X_i)}{1-\hat e(X_i)}\hat\mu_0(X_i)\hat S_0(k\mid X_i)\right\}.
\end{eqnarray*}
It has a similar mathematical form to the classic augmented inverse probability weighting estimator for the average treatment effect \citep{bang2005doubly}. We provide the properties of $\hat\tau^{\textup{aug}}$ in the following Proposition~\ref{prop::tau_aug}.

\begin{proposition}[Double robustness of $\hat\tau^{\textup{aug}}$]
\label{prop::tau_aug}
Suppose Assumptions~\ref{assump::randomization} and~\ref{assump::car} hold, and assume that $G_a(k\mid X)$ is correctly specified for $a=0,1$. $\hat\tau^{\textup{aug}}$ is a consistent estimator for $\tau$ if either $e(X)$ is correct, or both $\mu_a(X)$ and $S_a(k\mid X)$ are correct for $a=0,1$.
\end{proposition}

The augmented estimator $\hat\tau^{\textup{aug}}$ can be viewed as an intermediate bridge between the basic weighting estimator and the EIF-based estimator developed in the next section. Relative to $\hat\tau^{\textup{ipw}}$, it improves robustness by incorporating the outcome model, but it still requires correct specification of the censoring model $G_a(k\mid X)$. We provide additional motivations and remarks in Section~\ref{sec::supp_aug_estimator} of the Supplementary Material.

\section{A semiparametrically efficient and doubly robust estimator}
\label{sec::eif}

The augmented weighting estimator $\hat\tau^{\textup{aug}}$ improves robustness of $\hat\tau^{\textup{ipw}}$. However, it is not robust to the misspecification of the survival function for effect-uninformative IE, nor does it achieve the semiparametric efficiency bound. In this section, we show that $\hat\tau^{\textup{aug}}$ can be further improved. We derive the semiparametric efficient influence function (EIF) for $\tau$ to learn the best asymptotic efficiency a consistent estimator of $\tau$ can achieve and propose an asymptotically efficient and doubly robust estimator based on the EIF.

We first describe the full and observed data structures in our setting and introduce some additional notation. Ideally, we want to observe the full data $(X, Y^\c(1), Y^\c(0))$. The missingness of the full data comes from two strings. First, for a given treatment arm $a=0,1$, the effect-uninformative IE time $C(a)$ censors the full data because $Y^\c(a)$ is only observable if $C(a)\geq \{T(a)\wedge k\}$. Within each treatment group, we do not observe the full data due to censoring and only observe $$(\Delta(a)=1\{C(a)\geq T(a)\wedge k\}, \tilde{T}(a)=C(a)\wedge T(a)\wedge k, \Delta(a)Y^\c(a)).$$ Second, the treatment assignment generates another level of missingness, because, for each observation, we never simultaneously observe both composite potential outcomes $\{Y^\c(1), Y^\c(0)\}$ even without censoring. Therefore, the observed data is $$O = (A, \Delta = \Delta(A), \tilde{T} = \tilde{T}(A), \Delta Y^\c = \Delta(A)Y^\c(A)).$$

\subsection{EIF and EIF-based estimator}
In the following theorem, we provide the EIF for $\tau$.
\begin{theorem}[EIF for $\tau$]
\label{thm::eif_tau}
Under the nonparametric model, the EIF for $\mu_1$ is 
\begin{eqnarray}
    D_{1}(O) &=& \frac{A}{e(X)}\left\{ \frac{Y1(T\wedge C>k)}{G_1(k\mid X)} + \mu_1(X)S_1(k\mid X) \int_0^{\tilde{T}} \frac{\d M_{G_1}(t)}{S_1(t\mid X)G_1(t\mid X)} \right\} \notag \\
    && - \frac{A-e(X)}{e(X)}\mu_1(X)S_1(k\mid X) - \mu_1 \label{eqn::eif_mu1},
\end{eqnarray}
the EIF for $\mu_0$ is
\begin{eqnarray}
    D_{0}(O) &=& \frac{1-A}{1-e(X)}\left\{ \frac{Y1(T\wedge C>k)}{G_0(k\mid X)} + \mu_0(X)S_0(k\mid X) \int_0^{\tilde{T}} \frac{\d M_{G_0}(t)}{S_0(t\mid X)G_0(t\mid X)} \right\} \notag \\
    && - \frac{e(X)-A}{1-e(X)}\mu_0(X)S_0(k\mid X) - \mu_0 \label{eqn::eif_mu0},
\end{eqnarray}
and thus, the EIF for $\tau$ is $D_{\tau}(O) = D_1(O) - D_0(O)$, where $\d M_{G_a}(t) = 1(C\in \d t, \Delta = 0) - 1(\tilde{T}\geq t)\d\Lambda_{a}(t\mid X)$ with $\Lambda_{a}(t\mid X)$ denoting the conditional cumulative hazard function for the effect-uninformative IE $C$ in the treatment group $A=a$ for $a=0,1$. 
\end{theorem}

The $M_{G_a}(t)$ in the EIF is the martingale constructed from the censoring counting process. Intuitively, $1(C\in \d t, \Delta = 0)$ is the actual observed increment in the censoring counting process at time $t$, which records whether a censoring event has occurred, while $1(\tilde{T}\geq t)\d\Lambda_{a}(t\mid X)$ represents the expected increment in the counting process, given the history up to time $t$. The martingale $M_{G_a}(t)$ captures the difference between the actual observed events and their expected occurrences. The EIF implies another identification formula for $\tau$ by the property that $E\{D_{\tau}(O)\}=0$. Rearranging terms, we have
\begin{eqnarray}
    \tau &=& E\left\{\frac{AY1(T\wedge C>k)}{e(X)G_1(k\mid X)} - \frac{A-e(X)}{e(X)}\mu_1(X)S_1(k\mid X)\right\} \notag \\
    &&- E\left[\frac{(1-A)Y1(T\wedge C>k)}{\left\{1-e(X)\right\}G_0(k\mid X)} - \frac{e(X)-A}{1-e(X)}\mu_0(X)S_0(k\mid X) \right] \notag \\
    &&+ E\left[\frac{A}{e(X)}\mu_1(X)S_1(k\mid X) \int_0^{\tilde{T}} \frac{\d M_{G_1}(t)}{S_1(t\mid X)G_1(t\mid X)}\right] \notag \\
    &&- E\left[\frac{1-A}{1-e(X)} \mu_0(X)S_0(k\mid X) \int_0^{\tilde{T}} \frac{\d M_{G_0}(t)}{S_0(t\mid X)G_0(t\mid X)}\right] \label{eqn::eif_aug_term},
\end{eqnarray}
where the first two lines are the same as in the augmented weighting identification formula. Intuitively, the last two terms in~\eqref{eqn::eif_aug_term} provide additional correction to achieve robustness to misspecification of the survival function for effect-uninformative IE. These augmentation terms are zero at the population level under true $G_{a}(t\mid X)$. However, their empirical counterparts may deviate significantly from zero, providing diagnostic insight into possible misspecification of $G_{a}(t\mid X)$. The martingale term in Theorem~\ref{thm::eif_tau} is closely related to EIFs for right-censored survival data, such as \citet{westling2024inference}, because the nuisance tangent space for the censoring mechanism $C\mid A,X$ is the same as in standard censoring settings. The main difference is that our target is a treatment effect on a primary endpoint at a fixed horizon under a hybrid estimand with both effect-uninformative and effect-informative IEs, which introduces additional nuisance components, including the outcome regression and the survival model for the effect-informative IE.

The EIFs in Theorem~\ref{thm::eif_tau} motivate the following estimator for $\tau$,
\begin{eqnarray*}
    \hat{\tau}^{\textup{eif}} &=& \hat{\tau}^{\textup{aug}} + n^{-1}\sumn \frac{A_i}{\hat{e}(X_i)} \hat{\mu}_1(X_i)\hat{S}_1(k\mid X_i) \int_{0}^{\tilde{T}_i} \frac{\d M_{\hat{G}_1}(t)}{\hat{S}_1(t\mid X_i)\hat{G}_1(t\mid X_i)} \\
    &&- n^{-1}\sumn \frac{1-A_i}{1-\hat{e}(X_i)}\hat{\mu}_0(X_i)\hat{S}_0(k\mid X_i) \int_{0}^{\tilde{T}_i} \frac{\d M_{\hat{G}_0}(t)}{\hat{S}_0(t\mid X_i)\hat{G}_0(t\mid X_i)}.
\end{eqnarray*}
To gain insights, consider the first augmentation term in $\hat{\tau}^{\textup{eif}}$. Since the observed time points are discrete, the integral for observation $i$ can be estimated as follows:
\begin{eqnarray}
    \int_{0}^{\tilde{T}_i} \frac{\d M_{\hat{G}_1}(t)}{\hat{S}_1(t\mid X_i)\hat{G}_1(t\mid X_i)}
    &=& \sum_{t\leq \tilde{T}_i} \frac{1(\Delta_i = 0, C_i = t) - \hat{\lambda}_{1}(t\mid X_i)}{\hat{S}_1(t\mid X_i)\hat{G}_1(t\mid X_i)} \notag \\
    &=& - \sum_{t\leq \tilde{T}_i} \frac{\hat{\lambda}_{1}(t\mid X_i)}{\hat{S}_1(t\mid X_i)\hat{G}_1(t\mid X_i)} + \frac{1(\Delta_i = 0)}{\hat{S}_1(\tilde{T}_i\mid X_i)\hat{G}_1(\tilde{T}_i\mid X_i)}, \label{eqn::eif_est_form}
\end{eqnarray}
where $\hat{\lambda}_{1}(t\mid X_i)$ denotes the estimated conditional hazard function for the effect-uninformative IE at time $t$ given covariates $X_i$. In~\eqref{eqn::eif_est_form}, the first term is a sum of the ratio $-\hat{\lambda}_{1}(t\mid X_i) / \{\hat{S}_1(t\mid X_i)\hat{G}_1(t\mid X_i)\}$, where the summation is over all observed event time points before $\tilde{T}_i$. The second term is 0 for individuals who are not right-censored by the effect-uninformative IE, and is $1/\{\hat{S}_1(\tilde{T}_i\mid X_i)\hat{G}_1(\tilde{T}_i\mid X_i)\}$ for observations with an effect-uninformative IE occurred at time $\tilde{T}_i$.

To construct the estimator $\hat{\tau}^{\textup{eif}}$, we need to estimate the following nuisance parameters: the propensity score model $e(X)$, the outcome model $\mu_a(X)$ for $a=0,1$, and the survival functions for effect-informative and effect-uninformative IEs $S_a(t\mid X)$ and $G_a(t\mid X)$, respectively, for $t\leq k$ and $a=0,1$. Importantly, as shown in Theorem~\ref{thm::double_robustness}, the consistency of $\hat\tau^{\textup{eif}}$ does not require correct specification of all four nuisance parameters.

\subsection{Asymptotic properties of the EIF-based estimator}
We next discuss the asymptotic properties of $\hat{\tau}^{\textup{eif}}$. We first introduce additional notation. For $t\leq k$ and $a=0,1$, let $e^{*}$, $G_a^{*}$, $\mu_a^{*}$, and $S_a^{*}$ denote the probability limit of the estimated nuisance functions $\hat e$, $\hat G_a$, $\hat \mu_a$, $\hat S_a$, respectively, i.e., $\norm{\hat e - e^{*}} = o_P(1)$, $\norm{\hat G_a - G_a^{*}}=o_P(1)$, $\norm{\hat \mu_a - \mu_a^{*}}=o_P(1)$, and $\norm{\hat S_a - S_a^{*}}=o_P(1)$. If a given nuisance model is correctly specified, its corresponding limit equals the true function. For example, if the propensity score model is consistent, then $e^* = e$, and similar results hold for the other three nuisance components.

In the following theorem, we provide the double robustness of the EIF-based estimator $\hat{\tau}^{\textup{eif}}$. 

\begin{theorem}[Double robustness of $\hat{\tau}^{\textup{eif}}$]
\label{thm::double_robustness}
Under Assumptions~\ref{assump::randomization} and~\ref{assump::car}, $\hat\tau^{\textup{eif}}$ is doubly robust in the sense that it is consistent for $\tau$ if either $\{e^{*}(X)=e(X), G_a^{*}(t\mid X)=G_a(t\mid X)\}$ or $\{\mu_a^{*}(X)=\mu_a(X), S_a^{*}(t\mid X)=S_a(t\mid X)\}$ for $t\leq k$ and $a=0,1$.
\end{theorem}

Theorem~\ref{thm::double_robustness} shows that $\hat\tau^{\textup{eif}}$ is consistent if at least one of the following two sets of nuisance parameters are consistently estimated: (1) the outcome model $\mu_a(X)$ and the survival function for the effect-informative IE $S_a(t\mid X)$; (2) the propensity score model $e(X)$ and the survival function for the effect-uninformative IE $G_a(t\mid X)$, for $t\leq k$ and $a=0,1$. When our data is from an RCT, i.e., the propensity score $e(X)$ is known by design, so $\hat e(X)$ can be correctly specified. Consequently, the EIF-based estimator $\hat{\tau}^{\textup{eif}}$ is a consistent estimator for $\tau$ if either $G_a(t\mid X)$ is consistently estimated, or both $\mu_a(X)$ and $S_a(t\mid X)$ are consistently estimated, for $t\leq k$ and $a=0,1$.

To conduct statistical inference, we provide the asymptotic distribution of $\hat\tau^{\textup{eif}}$. We first introduce three technical conditions and then state the asymptotic result in a theorem.

\begin{assumption}[Consistency of the nuisance parameters]
\label{assump::consistency_nuisance}
Assume that either $\{e^{*}(X)=e(X), G_a^{*}(t\mid X)=G_a(t\mid X)\}$ or $\{\mu_a^{*}(X)=\mu_a(X), S_a^{*}(t\mid X)=S_a(t\mid X)\}$ for $t\leq k$ and $a=0,1$ is satisfied.
\end{assumption}

\begin{assumption}[Donsker Condition]
\label{assump::donsker}
The class of functions $\{(e,G_a,\mu_a,S_a):||e-e^*||<\delta, ||G_a-G_a^*||<\delta, ||\mu_a-\mu_a^*||<\delta, ||S_a-S_a^*||<\delta\}$ is Donsker for some $\delta>0$.
\end{assumption}

\begin{assumption}[Convergence rates of nuisance parameters]
\label{assump::rate_nuisance}
The convergence rate of the nuisance parameters estimation satisfies
$\{||\hat e - e^*|| + ||\hat G_a - G_a^*||\}\{||\hat \mu_a - \mu_a^*|| + ||\hat S_a - S_a^*||\} = o_P(n^{-1/2}) $ for $a=0,1$.
\end{assumption}

Assumption~\ref{assump::consistency_nuisance} requires either the propensity score and survival function of the effect-uninformative IE or the outcome and survival function of the effect-informative IE to be consistently estimated. It guarantees the consistency of $\hat{\tau}^{\textup{eif}}$. Assumption~\ref{assump::donsker} imposes restrictions on the nuisance model complexity and is a standard regularity condition \citep{van2000asymptotic}. We can also employ flexible machine learning models with cross-fitting techniques in the estimation of nuisance parameters to relax the Donsker condition \citep{pfanzagl1985contributions,klaassen1987consistent,zheng2011cross,chernozhukov2018double}. We provide a more detailed procedure in Section~\ref{sec::supp_crossfit} of the Supplementary Material, leaving a full theory and empirical study of implementations to future work. Finally, Assumption~\ref{assump::rate_nuisance} imposes additional restrictions on the rate of convergence of the nuisance parameters, in addition to all of them being consistently estimated. 

\begin{theorem}[Asymptotic distribution]
\label{thm::eif_asym_dist}
Under Assumptions~\ref{assump::randomization}--~\ref{assump::rate_nuisance}, the EIF-based estimator satisfies
\begin{equation*}
    n^{1/2} (\hat\tau^{\textup{eif}} - \tau) \ =\ n^{-1/2}\sumn D_{\tau}(O_i) + o_P(1).
\end{equation*}
The estimator $\hat\tau^{\textup{eif}}$ is a consistent and asymptotically Normal estimator of $\tau$ with asymptotic variance equal to $E\{D_{\tau}^2(O)\}$, thus achieving the semiparametric efficiency bound.
\end{theorem}

Therefore, we can construct the variance estimator based on the semiparametric efficiency bound by taking the empirical analog of the plug-in estimation $n^{-1}\sumn \hat D_{\tau}^2(O_i)$. We can also use a nonparametric bootstrap to estimate variance and conduct statistical inference if the nuisance estimators satisfy certain smoothness conditions. 

\subsection{Comparing the estimators}
\label{sec::comparison_of_estimators}

In real data analysis, we recommend implementing and comparing all four proposed estimators to assess whether they provide coherent scientific implications. Theoretical comparisons among these estimators primarily focus on two key dimensions: robustness to nuisance model misspecification and asymptotic efficiency. In practice, examining how the point estimates differ across these estimators can provide insight into the presence of nuisance model misspecification. Specifically, certain pairwise differences among the estimators can serve as informal diagnostics. For example, if $e(X)$ is correctly specified, the difference between $\hat\tau^{\textup{ipw}}$ and $\hat\tau^{\textup{aug}}$ should converge to zero. Similarly, if $\mu_a(X)$ and $S_a(k\mid X)$ are correctly specified, the difference between $\hat\tau^{\textup{out}}$ and $\hat\tau^{\textup{aug}}$ should converge to zero. Finally, if $G_a(t\mid X)$ is correctly specified, the difference between $\hat\tau^{\textup{eif}}$ and $\hat\tau^{\textup{aug}}$ should converge to zero. Therefore, when implementing all four estimators in practice, the observed difference among them may serve as an informal diagnostic tool to detect possible nuisance model misspecification. We provide more details in Section~\ref{sec::supp_pairwise_diff} in the supplementary material.

Among the four estimators, a key comparison lies between the EIF-based estimator $\hat\tau^{\textup{eif}}$ with the augmented weighting estimator $\hat\tau^{\textup{aug}}$. On the one hand, $\hat\tau^{\textup{eif}}$ is more robust and asymptotically efficient: it allows for misspecification of the censoring model $C\mid A,X$ and it achieves the semiparametric efficiency bound under correct nuisance specification. Continuing the intuition discussed in Section~\ref{subsec::ps_aug_est}, $\hat\tau^{\textup{aug}}$ only orthogonalizes the part corresponding to $A\mid X$ but not $C\mid X,A$, making it robust only to misspecification of the $A\mid X$ model, not the $C\mid X,A$ model. In contrast, $\hat\tau^{\textup{eif}}$ accounts for both possible sources of misspecification, leading to greater robustness and efficiency. On the other hand, $\hat\tau^{\textup{aug}}$ is more straightforward to implement in practice, as it only requires the estimation of $S_a(k\mid X)$ and $G_a(k\mid X)$ at time $k$ for $a=0,1$, while $\hat\tau^{\textup{eif}}$ requires consistent estimation of the entire survival curves $S_a(t\mid X)$ and $G_a(t\mid X)$ for $t\leq k$ and $a=0,1$. We consider the choice between $\hat\tau^{\textup{eif}}$ and $\hat\tau^{\textup{aug}}$ as a trade-off between implementation simplicity and robustness and efficiency of the estimators.

Finally, we note the theoretical possibility of applying a martingale-based correction to the basic weighting estimator $\hat{\tau}^{\textup{ipw}}$, in a manner analogous to the correction in the EIF-based estimator $\hat\tau^{\textup{eif}}$ compared with $\hat\tau^{\textup{aug}}$. Such a corrected estimator would be robust to misspecification of $G_a(t\mid X)$. However, it remains less attractive in practice: it is neither robust to misspecification of the propensity score nor asymptotically efficient, and it lacks the implementation simplicity of $\hat\tau^{\textup{aug}}$. For these reasons, we do not recommend it, though we include its explicit form in Section~\ref{sec::supp_martingale_correction_form} of the supplementary material only for theoretical completeness.

\subsection{Simulation results}
\label{sec::simulation_brief_in_main}
We conduct Monte Carlo simulations to evaluate the finite-sample performance of four estimators under various combinations of correctly specified and misspecified nuisance models. Consistent with our theory, $\hat\tau^{\textup{eif}}$ demonstrates the greatest robustness across all misspecification scenarios, maintaining low bias and valid coverage even when multiple nuisance models are incorrect. We relegate the detailed results to Section~\ref{sec::simulation} in the supplementary material.

\section{Analyzing two phase-3 trials}
\label{sec::application}

\subsection{Data analysis}
In this section, we re-analyze data from two trials on systemic lupus erythematosus \citep{morand2023baricitinib, petri2023baricitinib}. Both trials are double-blinded, randomized, placebo-controlled phase-3 trials. There are three treatment arms: 2mg baritinib, 4mg baritinib, and placebo. The primary outcome in both studies is the Systemic lupus erythematosus Responder Index 4 (SRI4) at week 52, a binary composite endpoint that reflects clinical response. A participant is classified as a responder if they show a predefined improvement in disease activity, without overall worsening or the emergence of significant disease activity in new organ systems. Baseline covariates used for model fitting include geographic region, baseline corticosteroid use, and the Physician’s Global Assessment score measured at baseline. We provide detailed baseline characteristics in Section~\ref{sec::supp_eda} of the Supplementary Material. Although the two trials each have three treatment arms, our current framework is formulated for binary treatment contrasts. Accordingly, in the data analysis we conduct the two standard pairwise comparisons against placebo, namely 2mg versus placebo and 4mg versus placebo. A fully joint multi-arm formulation is a natural extension, but is beyond the scope of the current paper.

A substantial proportion of participants in both trials experienced IEs: 218 out of 760 in one trial and 211 out of 775 in the other. As described in Section~\ref{sec::proposal_and_contribution}, we classify these IEs into two categories: effect-informative IEs (82.6\% and 84.4\% in the two trials, respectively) and effect-uninformative IEs (17.4\% and 15.6\%). We conduct separate comparisons of the 2mg and 4mg baricitinib treatment arms versus the placebo group, yielding four pairwise analyses with sample sizes ranging from 505 to 517. Among participants who experienced an effect-informative IE, the median time to IE ranges from 16 to 23 weeks. Effect-uninformative IEs are sparse (3--8\% across arms) and tend to occur slightly later (median 18--25 weeks). We present the Kaplan--Meier-based cumulative incidence of each IE type by arm and trial in Section~\ref{sec::supp_eda} of the Supplementary Material. 

For each comparison, we apply all four estimators: $\hat\tau^{\textup{out}}$, $\hat\tau^{\textup{ipw}}$, $\hat\tau^{\textup{aug}}$, and $\hat\tau^{\textup{eif}}$. We use logistic regression to estimate the propensity score and outcome models, and use Cox proportional hazards regression to separately estimate the survival functions for the two types of IEs. We conduct variance estimation and inference using a nonparametric bootstrap with 500 replicates. Table~\ref{tab::application_result} presents the resulting point estimates, bootstrap standard errors, $p$-values, and 95\% confidence intervals. We did not apply a formal multiplicity adjustment in this data analysis because the comparisons are presented primarily for descriptive and methodological illustration rather than as a confirmatory family of hypothesis tests.

\begin{table}[h]
\caption{Treatment effect of Baricitinib on the primary outcome SRI4. The first four columns report the results of Trial 1 and the last four columns report those of Trial 2. Within each trial, the first five rows report the point estimate, bootstrap standard error, $p$-value, and lower and upper 95\% confidence limits for each estimator of the 2mg treatment arm, and the last five rows report those of the 4mg treatment arm. For each trial and each treatment arm, we report the results based on all four estimators.}
\doublespacing
\centering
\begin{small}
\begin{tabular}{ccccccccccc}
\\
\hline
\hline
 & & \multicolumn{4}{c}{Trial 1 \citep{petri2023baricitinib}} && \multicolumn{4}{c}{Trial 2 \citep{morand2023baricitinib}}\\
         \cline{3-6} \cline{8-11} 
         &  &  $\hat{\tau}^{\textup{out}}$&  $\hat{\tau}^{\textup{ipw}}$& $\hat{\tau}^{\textup{aug}}$& $\hat{\tau}^{\textup{eif}}$  && $\hat{\tau}^{\textup{out}}$& $\hat{\tau}^{\textup{ipw}}$&$\hat{\tau}^{\textup{aug}}$& $\hat{\tau}^{\textup{eif}}$ \\
         \hline
         2mg&  point&  0.030&  0.029&  0.026&  0.026&  &  0.019&  0.019&  0.022& 0.022
\\
         &  se&  0.042&  0.043&  0.042&  0.043&  &  0.042&  0.043&  0.043& 0.042
\\
         &  $p$-value&  0.479&  0.504&  0.534&  0.540&  &  0.643&  0.662&  0.602& 0.606
\\
         &  CI lower&  $-0.053$&  $-0.055$&  $-0.057$&  $-0.057$&  &  $-0.062$&  $-0.065$&  $-0.061$& $-0.061$
\\
         &  CI upper&  $0.113$&  $0.113$&  $0.110$&  $0.110$&  &  $0.101$&  $0.102$&  $0.106$& $0.104$
\\
\hline
         4mg&  point&  0.113&  0.120&  0.115&  0.113&  &  $-0.002$&  $-0.002$&  $-0.002$& $-0.002$
\\
         &  se&  0.046&  0.046&  0.046&  0.046&  &  0.042&  0.042&  0.042& 0.042
\\
         &  $p$-value&  0.013&  0.008&  0.012&  0.013&  &  0.961&  0.962&  0.966& 0.969
\\
         &  CI lower&  $0.023$&  $0.030$&  $0.025$&  $0.023$&  &  $-0.084$&  $-0.084$&  $-0.084$& $-0.084$
\\
         &  CI upper&  $0.204$&  $0.209$&  $0.204$&  $0.203$&  &  $0.080$&  $0.080$&  $0.081$& $0.080$
\\
\hline
    \end{tabular}
\end{small}
\label{tab::application_result}
\end{table}

\subsection{Interpretation of the statistical results}

Across both trials and treatment arms, the point estimates and $p$-values from the four methods are generally consistent. Our results suggest a statistically significant positive effect of 4mg baricitinib on the SRI4 outcome in Trial 1, while in Trial 2, the effect is not significant and the point estimate is slightly negative. No significant treatment effect is observed for the 2 mg baricitinib dose in either trial. As discussed in Section~\ref{sec::comparison_of_estimators}, though not a formal statistical test, qualitatively, the robustness of estimates across the four estimators in Table~\ref{tab::application_result} provides empirical evidence that our nuisance models may not be severely misspecified.

In these two immunology trials, our estimand has a direct clinical interpretation. For each baricitinib versus placebo comparison, it targets the difference in the probability of week-52 SRI4 response under a hybrid strategy for handling IEs. If a patient experiences an effect-informative IE before week 52, such as discontinuation due to adverse events or lack of efficacy, that patient is counted as a non-responder under the composite strategy. If a patient experiences an effect-uninformative IE, such as relocation, a pandemic-related disruption, or sponsor-driven site closure, the estimand instead asks what the week-52 SRI4 outcome would have been had that external disruption not occurred. Thus, the estimand addresses the following clinically relevant question: \textit{What is the treatment effect on week-52 SRI4 if external disruptions had not occurred, while still counting clinically meaningful treatment failures as failures?} For example, the estimated effect of about 0.11 for the 4mg arm in Trial 1 implies that, under this hybrid strategy, 4mg baricitinib increases the probability of week-52 SRI4 response by approximately 11\% points relative to placebo.

\subsection{Comparison with ad-hoc methods}

Next, we compare our proposed methods with two commonly used ad-hoc methods. The first method is the \textit{non-responder rule} (often referred to as ``non-responder imputation'' in the applied literature). We adopt the term ``rule'' instead of ``imputation'' to emphasize that this approach redefines the endpoint to take the value $0$ upon an IE, rather than treating the outcome as missing and imputing it. It is equivalent to using the composite outcome strategy that treats all IEs as effect-informative and assigns the failure value $0$ upon any IE. The second method naively applies a hypothetical strategy to all IEs, assuming they are independent of potential outcomes conditional on observed covariates. 

Table~\ref{tab::application_adhoc} reports the estimated treatment effects of both treatment arms across the two trials using the inverse probability weighting estimator within each approach. We would like to emphasize that this is \textit{not} merely a comparison of different estimators for a common estimand. Instead, the three approaches correspond to different choices of estimand because they classify and handle IEs in different ways. Therefore, the results illustrate how the estimated treatment effect changes when different IE classifications, estimand strategies, and identifying assumptions are adopted.

The non-responder rule method generally yields slightly smaller effect estimates than our proposed methods, although the differences are modest, likely due to the relatively low proportion of effect-uninformative IEs in both trials. Nevertheless, our proposed approach targets a clinically more meaningful causal estimand.
The second ad hoc method naively applies the hypothetical strategy to all IEs, assuming that their occurrence is conditionally independent of the potential outcomes given baseline covariates. This assumption is untestable and likely violated in practice. For example, treatment discontinuation due to adverse events or lack of efficacy is plausibly related to a patient's potential outcome. As shown in Table~\ref{tab::application_adhoc}, this method produces treatment effect estimates that differ substantially from those obtained using our proposed approach, potentially leading to misleading clinical conclusions. These empirical findings are consistent with the large-sample bias observed in the simulation study and underscore the importance of using appropriate strategies tailored to different types of IEs.

\begin{table}[h]
\caption{Treatment effect of Baricitinib on the primary outcome SRI4 using non-responder rule (NRR) and hypothetical strategy (HS). The first two columns report the results of Trial 1 and the last two columns report those of Trial 2. The first five rows report the point estimate, bootstrap standard error, $p$-value, and lower and upper 95\% confidence limits for the 2mg treatment arm, and the last five rows report those of the 4mg treatment arm. For each trial and each treatment arm, we report the results based on the inverse probability-weighting estimator.}
\doublespacing
\centering
\begin{small}
\begin{tabular}{ccccccc}
\\
\hline
\hline
 & & \multicolumn{2}{c}{Trial 1} && \multicolumn{2}{c}{Trial 2 }\\
         \cline{3-4} \cline{6-7} 
         & & NRR& HS && NRR&HS
\\
\hline
 2mg& point& 0.039& 0.011 && 0.010&0.022
\\
 & se& 0.043& 0.049 && 0.042&0.050
\\
 & $p$-value& 0.362& 0.815 && 0.817&0.654
\\
 & CI lower& $-0.045$& $-0.085$ && $-0.072$&$-0.076$
\\
 & CI upper& $0.123$& $0.107$ && $0.092$&$0.120$
\\
\hline
 4mg& point& 0.093& 0.073 && $-0.008$&0.044
\\
 & se& 0.044& 0.050 && 0.041&0.051
\\
 & $p$-value& 0.036& 0.147 && 0.847&0.382
\\
 & CI lower& $0.007$& $-0.025$ && $-0.088$&$-0.056$
\\
 & CI upper& $0.179$& $0.171$ && $0.072$&$0.144$
\\
\hline
\end{tabular}
\end{small}
\label{tab::application_adhoc}
\end{table}

\subsection{Robustness checks}
\label{sec::robustness_checks}

As a robustness check to the proportional hazards assumption embedded in the Cox nuisance models, we replaced all four Cox models with log-normal accelerated failure time (AFT) models, which impose no proportional hazards constraint. The AFT survival curves are evaluated analytically at all observed follow-up times and fed into the same martingale integral computation as the Cox-based EIF estimator. The AFT sensitivity results are nearly identical to the Cox-based results across all four comparisons: for the 4mg arm in Trial 1, the AFT EIF estimate is 0.113 (95\% CI: [0.023, 0.203]), identical to the Cox-based result. Full results are reported in Section~\ref{sec::supp_aft} of the Supplementary Material. This close agreement suggests that the conclusions are robust to the choice of parametric survival model for the nuisance functions.

As a further robustness check on the IE classification and on how potential misspecifications affect the estimands and estimates, we reclassified IE categories under alternative clinically possible rules and re-estimated the treatment effects. This robustness check probes the sensitivity of the results to the classification decisions. Under all scenarios we tested, the estimated treatment effects and their confidence intervals remain close to the primary analysis results, and the statistical conclusion for the 4mg baricitinib arm in Trial~1 is preserved. Full results are reported in Section~\ref{sec::robustness_ie_class} of the Supplementary Material.

\subsection{Beyond $\tau$: causal risk ratio and odds ratio}
For binary composite outcomes, our framework also extends directly to other effect measures, such as the causal risk ratio and odds ratio, by first estimating the arm-specific mean composite outcomes and then applying the corresponding smooth transformation. The associated EIFs follow from those for the arm-specific means via the delta method. Although mainly focusing on the risk difference, $\tau$, in the main paper, we also implement causal risk ratio and odds ratio estimates for all four comparisons and report the full results in Section~\ref{sec::supp_rr_or} of the Supplementary Material.

\section{Future directions}
\label{sec::discussion}

In this study, we address the challenges posed by IEs in RCTs by proposing methods to handle competing IEs. We classify IEs into effect-informative and effect-uninformative events and apply different strategies to identify a clinically meaningful causal effect. For effect-informative IEs, which are often informative about a patient’s outcome, we use a composite variable strategy that assigns an outcome value indicative of treatment failure. For effect-uninformative IEs, we apply a hypothetical strategy, assuming their timing is conditionally independent of the outcome given treatment and baseline covariates, and envisioning a scenario in which such events do not occur. The central thesis of this paper is to address the challenge of competing IEs, where the first IE censors all subsequent ones. In this paper, we propose a principled framework that carefully formulates the estimand, establishes its nonparametric identification and semiparametric estimation theory, and introduces weighting, outcome regression, and doubly robust estimators. While our proposed framework provides a rigorous and flexible approach for handling competing IEs in RCTs, several challenges and extensions remain open for future research.

\subsection{Data collection and IE classification}
Our proposed methods have broad applicability across various therapeutic areas, including immunology, oncology, and cardiology, where treatment discontinuation and other IEs frequently occur. However, the effectiveness of these approaches relies on the accurate classification of IEs, which should be performed in collaboration with clinicians to ensure clinical relevance in trial analyses. To support this, a modernized case report form is needed to enable more granular documentation of the timing, reasons, and magnitude of treatment discontinuation. For example, the reason for the event should be specified, such as discontinuation due to toxicity versus lack of efficacy. In some cases, the event may need to meet a threshold of magnitude, such as the use of additional medication exceeding a specified duration or dose. Additionally, the timing of the event may be relevant, particularly in relation to its proximity to outcome assessment. Fortunately, a cross-industry PHUSE working group is tackling this problem \cite{phuse2025estimands}.

\subsection{Random assessment time $K$}
In practice, the measurement of the outcome of interest $Y$ does not necessarily happen at a fixed point $k$, but within a prespecified assessment window. For example, a protocol may define the primary endpoint at week 52 while allowing the final study visit to occur within a window such as weeks 50--54. In that case, the actual visit time at which the outcome is measured varies across patients. Let $K$ denote this realized assessment time. The outcome remains the same clinical endpoint of interest, but it is recorded at the patient-specific visit time $K$ rather than at a common fixed time $k$. If this variability in visit timing is driven by protocol logistics or other external factors that are unrelated to treatment assignment, potential outcomes, and IEs, then $K$ can be treated as a random variable independent of treatment $A$, potential outcomes $Y(a,c=\infty)$, and both $T(a)$ and $C(a)$ for $a=0,1$. Under this extension, the analogous versions of the identification formulas~\eqref{eqn::id_out} and \eqref{eqn::id_ipw} are obtained by replacing $k$ with $K$, i.e.,
\begin{eqnarray}
    \tau &=& E\left\{ \mu_1(X) S_1(K\mid X) - \mu_0(X) S_0(K\mid X) \right\} \label{eqn::id_out_randomK}\\
     &=& E\left[ \frac{AY1(T\wedge C>K)}{e(X)G_1(K\mid X)} - \frac{(1-A)Y1(T\wedge C>K)}{\left\{1-e(X)\right\}G_0(K\mid X)} \right]. \label{eqn::id_ipw_randomK}
\end{eqnarray}

If $K$ is a pre-treatment covariate that is observed, the proposed four estimators carry over by replacing the fixed $k$ with the observed values of $K$. In contrast, if $K$ is itself affected and is a post-treatment variable, we can construct a weighting estimator following~\eqref{eqn::id_ipw_randomK}. However, \eqref{eqn::id_out_randomK} no longer provides a feasible identification formula unless $S_a(K\mid X)$ is identified for $a=0,1$. Considering the presence of three competing time-to-event random variables, it becomes necessary to employ competing risks models. We defer this analysis to future work.


\section*{Acknowledgement}
Sizhu Lu and Peng Ding were partially supported by the U.S. National Science Foundation grants \#1745640 and \#1945136. Ting Ye was partially supported by the HIV Prevention Trials Network (HPTN) and NIH grant: NIAID 5 UM1 AI068617.

\section*{Supplementary material}
The supplementary material includes additional technical details, simulation results, and proofs of all theorems and propositions.

\bibliographystyle{apalike}
\bibliography{missing_bib}

\newpage

\appendix

\newpage
\begin{appendix}
\begin{center}
  \LARGE {\bf Supplementary Material}
\end{center}

\pagenumbering{arabic} 
\renewcommand*{\thepage}{S\arabic{page}}

\setcounter{equation}{0} 
\global\long\def\theequation{S\arabic{equation}}%
 \setcounter{assumption}{0} 
\global\long\def\theassumption{S\arabic{assumption}}%
 \setcounter{theorem}{0} 
\global\long\def\thetheorem{S\arabic{theorem}}%
 \setcounter{proposition}{0} 
\global\long\def\theproposition{S\arabic{proposition}}%
 \setcounter{definition}{0} 
\global\long\def\thedefinition{S\arabic{definition}}%
 \setcounter{example}{0} 
\global\long\def\theexample{S\arabic{example}}%
 \setcounter{figure}{0} 
\global\long\def\thefigure{S\arabic{figure}}%
 \setcounter{table}{0} 
\global\long\def\thetable{S\arabic{table}}%

Section~\ref{sec::additional_results} provides additional technical results on details of the augmented estimator $\hat{\tau}^{\textup{aug}}$, comparing the four proposed estimators, another martingale-corrected weighting estimator discussed in Section~\ref{sec::comparison_of_estimators} of the main paper, and a cross-fitted debiased machine learning implementation procedure of the EIF-based estimator.

Section~\ref{sec::supp_additional_data_analysis} provides additional data analysis results, including classification of all IEs in the two immunology trials, exploratory data analysis for baseline covariates and IEs, additional results on robustness check, estimation results targeting causal risk ratio and odds ratios, and remarks on computational time of our proposed estimators.

Section~\ref{sec::simulation} reports detailed results from Monte Carlo simulations that assess the finite-sample performance of our proposed estimators. 

Section~\ref{sec::proofs} provides proofs of all theorems and propositions.

\section{Additional technical results}
\label{sec::additional_results}

\subsection{Additional details on $\hat\tau^{\textup{aug}}$}
\label{sec::supp_aug_estimator}

In this subsection, we discuss additional motivations and remarks about the proposed estimator $\hat\tau^{\textup{aug}}$. First, by combining the two identification formulas~\eqref{eqn::id_out} and~\eqref{eqn::id_ipw}, we derive the following augmented identification formula:
\begin{eqnarray}
    \tau &=& E\left\{\frac{AY1(T\wedge C>k)}{e(X)G_1(k\mid X)} - \frac{A-e(X)}{e(X)}\mu_1(X)S_1(k\mid X)\right\}\notag \\
     &&- E\left[\frac{(1-A)Y1(T\wedge C>k)}{\left\{1-e(X)\right\}G_0(k\mid X)} - \frac{e(X)-A}{1-e(X)}\mu_0(X)S_0(k\mid X) \right] \label{eqn::id_aug_ipw}.    
\end{eqnarray}

Equation~\eqref{eqn::id_aug_ipw} can be interpreted both as a modified form of the outcome regression identification formula~\eqref{eqn::id_out}, and as an augmented form of the weighting identification formula in~\eqref{eqn::id_ipw}. Under the true model at the population level, the correction terms have mean $0$, and thus~\eqref{eqn::id_aug_ipw} holds by construction. However, it becomes meaningful in the presence of possible model misspecification, where the augmentation plays a crucial role in improving robustness.

Based on~\eqref{eqn::id_aug_ipw}, we construct the estimator $\hat\tau^{\textup{aug}}$ that augments the weighting estimator $\hat\tau^{\textup{ipw}}$ using the estimated outcome models. The improvement in its robustness compared with $\hat\tau^{\textup{ipw}}$ is summarized in Proposition~\ref{prop::tau_aug}, which shows that the consistency of $\hat\tau^{\textup{aug}}$ depends on correctly specifying $G_a(k \mid X)$ for $a = 0, 1$, while it remains robust to misspecification of the propensity score $e(X)$. To build intuition, from the perspective of semiparametric efficiency theory, Equation~\eqref{eqn::id_aug_ipw} can be viewed as a projection of the weighting identification formula~\eqref{eqn::id_ipw} onto the nuisance tangent space of the propensity score model $A\mid X$. However, because it is not further projected onto the nuisance tangent space of the $G_a(k\mid X)$ model, the resulting estimator does not retain robustness to misspecification of $G_a(k\mid X)$. We will further address the issue in Section~\ref{sec::eif}, where we introduce an alternative estimator that is robust to the misspecification of both models.

The augmented estimator $\hat\tau^{\textup{aug}}$ dominates the weighting estimator $\hat\tau^{\textup{ipw}}$ in terms of robustness. The consistency of $\hat\tau^{\textup{aug}}$ is contingent on less restrictive requirements on nuisance parameter estimation in the sense that $\hat\tau^{\textup{aug}}$ is consistent whenever $\hat\tau^{\textup{ipw}}$ is. Consistency of $\hat\tau^{\textup{aug}}$ requires the correct specification of the survival function for effect-uninformative IE $G_a(k\mid X)$ for $a=0,1$. Given that $G_a(k\mid X)$ is correctly modeled, $\hat\tau^{\textup{aug}}$ is doubly robust because it is consistent if either the propensity score model is correct, or the outcome regression and the survival model for effect-informative IEs are both correct. There is no clear dominance between $\hat\tau^{\textup{aug}}$ and $\hat\tau^{\textup{out}}$ as they require consistent estimation of different sets of nuisance parameters. For all three estimators $\hat\tau^{\textup{out}}$, $\hat\tau^{\textup{ipw}}$, and $\hat\tau^{\textup{aug}}$, we can construct variance estimators using nonparametric bootstrap.

\subsection{Pairwise differences between the estimators}
\label{sec::supp_pairwise_diff}
As discussed in Section~\ref{sec::comparison_of_estimators}, the observed differences among the four estimators provide insight into the presence of nuisance model misspecification. To illustrate the idea, consider the treated arm as an example. Let $\hat\mu_1^{\dagger}$ denote the treated arm counterpart in $\hat\tau^{\dagger}$ for $\dagger \in \{\textup{out, ipw, aug, eif\}}$, and let $\xrightarrow{\textup{p}}$ denote convergence in probability. The probability limits of the pairwise differences between these estimators are given by
\begin{eqnarray*}
    \hat\mu_1^{\textup{aug}}-\hat\mu_1^{\textup{ipw}} &\xrightarrow{\textup{p}}& E\left\{\frac{e^{*}(X)-e(X)}{e^{*}(X)}\mu_1^{*}(X)S_1^{*}(k\mid X)\right\}, \\
    \hat\mu_1^{\textup{aug}}-\hat\mu_1^{\textup{out}} &\xrightarrow{\textup{p}}& E\left[\frac{e(X)}{e^{*}(X)}\left\{\mu_1^{*}(X)S_1^{*}(k\mid X)-\mu_1(X)S_1(k\mid X)\right\}\right], \\
    \hat\mu_1^{\textup{eif}}-\hat\mu_1^{\textup{aug}} &\xrightarrow{\textup{p}}& E\left[\frac{e(X)}{e^{*}(X)}\mu_1^{*}(X)S_1^{*}(k\mid X) \int_{0}^{k} \frac{S_1(t\mid X) G_1(t\mid X)}{S_1^{*}(t\mid X)G_1^{*}(t\mid X)}\{\d\Lambda_{1}(t\mid X) - \d\Lambda_{1}^{*}(t\mid X)\}\right].
\end{eqnarray*}

These probability limits indicate the impact of nuisance model misspecification on the pairwise comparisons. For example, if $G_1(t\mid X)$ is correctly specified, the expected difference between $\hat\mu_1^{\textup{eif}}$ and $\hat\mu_1^{\textup{aug}}$ should be small. Similar logic applies to other comparisons. Therefore, when implementing all four estimators in practice, the observed difference among them may serve as an informal diagnostic tool to detect possible nuisance model misspecification. 

\subsection{The martingale-corrected weighting estimator}
\label{sec::supp_martingale_correction_form}
In this subsection, we present the explicit form of the martingale-corrected weighting estimator introduced in Section~\ref{sec::comparison_of_estimators}. Drawing from the proof of Theorem~\ref{thm::eif_tau} in Section~\ref{sec::proof_eif_tau}, if we do not project onto the propensity score nuisance tangent space in Step 2, the following identification formula for $\mu_1$ holds:
\begin{eqnarray*}
    \mu_1 &=& E\left[\frac{A}{e(X)}\left\{\frac{Y 1(T \wedge C>k)}{G_1(k \mid X)}+\mu_1(X) S_1(k \mid X) \int_0^{\tilde{T}} \frac{\mathrm{~d} M_{G_1}(t)}{S_1(t \mid X) G_1(t \mid X)}\right\}\right],
\end{eqnarray*}
which motivates the following estimator for $\mu_1$:
\begin{eqnarray*}
    \hat{\mu}_1^{\textup{mc-ipw}} &=& n^{-1}\sumn \frac{A_i Y_i 1(T_i \wedge C_i > k)}{\hat{e}(X_i)\hat{G}_1(k\mid X_i)} + n^{-1}\sumn \frac{A_i}{\hat{e}(X_i)} \hat{\mu}_1(X_i)\hat{S}_1(k\mid X_i) \int_{0}^{\tilde{T}_i} \frac{\d M_{\hat{G}_1}(t)}{\hat{S}_1(t\mid X_i)\hat{G}_1(t\mid X_i)}.
\end{eqnarray*}
The control counterpart, $\hat{\mu}_0^{\textup{mc-ipw}}$, is defined analogously, and the estimator for $\tau$ is given by $\hat{\tau}^{\textup{mc-ipw}} = \hat{\mu}_1^{\textup{mc-ipw}} - \hat{\mu}_0^{\textup{mc-ipw}}$. 

Under Assumuptions~\ref{assump::randomization} and~\ref{assump::car}, and assuming correct specification of $e(X)$, the estimator $\hat{\tau}^{\textup{mc-ipw}}$ is consistent for $\tau$ if either $G_a(t\mid X)$ is correct, or both $\mu_a(X)$ and $S_a(t\mid X)$ are correct for $t\leq k$ and $a=0,1$. Although the martingale correction enhances robustness to misspecification of $G_a(t\mid X)$, $\hat{\tau}^{\textup{mc-ipw}}$ is still less attractive than $\hat{\tau}^{\textup{eif}}$: it does not achieve the efficiency bound and is not robust to misspecification of $e(X)$. Moreover, it does not inherit the implementation simplicity of $\hat{\tau}^{\textup{aug}}$. In particular, it requires estimating the full survival curves for both the effect-uninformative IEs $G_a(t\mid X)$ and effect-informative IEs $S_a(t\mid X)$ over $t\leq k$, making it computationally demanding without offering clear practical advantages.

\subsection{A cross-fitted debiased machine learning implementation of the EIF-based estimator}
\label{sec::supp_crossfit}

The EIF-based estimator in Section~\ref{sec::eif} admits a natural cross-fitted implementation in the spirit of double/debiased machine learning \citep{chernozhukov2018double}. We briefly outline one possible algorithm here.

\begin{itemize}
    \item Step 1: Randomly partition the units into $L$ approximately equal-sized folds $\mathcal{I}_1,\ldots,\mathcal{I}_L$.
    \item Step 2: For each fold $\ell=1,\ldots,L$, use the units outside $\mathcal{I}_\ell$ to estimate the nuisance functions $e(X), \mu_a(X), S_a(t\mid X), G_a(t\mid X)$ for a=0,1. Denote the resulting estimators by $\hat e^{(-\ell)}$, $\hat\mu_a^{(-\ell)}$, $\hat S_a^{(-\ell)}$, and $\hat G_a^{(-\ell)}$.
    \item Step 3: For each unit $i\in\mathcal{I}_\ell$, evaluate the EIF using the nuisance estimators trained on the complement of fold $\ell$. That is, compute the cross-fitted version of the EIF contribution $\hat D_{\tau,i}^{(-\ell)}$ by substituting $\hat e^{(-\ell)}$, $\hat\mu_a^{(-\ell)}$, $\hat S_a^{(-\ell)}$, and $\hat G_a^{(-\ell)}$ into the expression of $D_\tau(O_i)$ in Theorem~\ref{thm::eif_tau}.
    \item Step 4: Define the cross-fitted EIF-based estimator by averaging the fold-specific scores: $\hat\tau^{\textup{cf}} = \frac{1}{n}\sum_{i=1}^n \hat\psi_i,$ where $\hat\psi_i$ denotes the plug-in estimating expression corresponding to unit $i$ with nuisance functions estimated out-of-fold. 
    \item Step 5: An asymptotic variance estimator can be constructed using bootstrap or the empirical variance of the cross-fitted EIF contributions, $n^{-1}\sum_{i=1}^n (\hat D_{\tau,i}^{(-\ell(i))})^2,$ where $\ell(i)$ is the fold containing unit $i$. Under regularity conditions, this yields a valid inference in the usual way.
\end{itemize}

This cross-fitted construction follows the same general principle as standard double/debiased machine learning: nuisance functions are estimated on auxiliary subsamples, while the orthogonal score is evaluated on held-out units. In our setting, the main additional practical burden is that the nuisance functions include not only the outcome regression and propensity score, but also the survival functions for the effect-informative and effect-uninformative IEs.

\section{Additional data analysis results}
\label{sec::supp_additional_data_analysis}

\subsection{Detailed classification of IEs in the immunology trials}
\label{sec::supp_ie_classification}

In the two immunology trials, IE categories were classified through manual review of the site-level documentation. We use the terms \emph{effect-informative} and \emph{effect-uninformative} to emphasize whether the event is judged to carry information about the patient’s underlying treatment response. Table~\ref{tab::supp_ie_classification} provides the complete list of IE categories observed in the two trials, together with their frequencies, assigned classifications, and the rationale used in our review.

Guided by the recommendations in the PHUSE white paper on estimand implementation and data standards \citep{phuse2025estimands}, we use the following general rule for classification. We classified an IE as effect-uninformative only when the documentation suggested that it arose from external or administrative reasons plausibly unrelated to the biological effect of treatment, such as relocation, pandemic-related disruption, or sponsor-driven site closure. We classified an IE as effect-informative when it was directly related to adverse events, lack of efficacy, patient health status, treatment management, or when the available documentation was too limited to justify an effect-uninformative classification. In ambiguous cases, we classified the IE as effect-informative to avoid overestimating arm-specific mean outcomes. Several categories required case-by-case review because the high-level label alone was not always sufficiently informative. In particular, categories such as personal or family reason, withdrawal by subject, protocol deviation, and unable to contact were reviewed using the site-level records. This review was used to determine whether the event was plausibly external to the patient's treatment response or whether it was more credible to regard the event as carrying information about efficacy, tolerability, or patient status.

\begin{longtable}{p{2cm}c c c p{8cm}}
\caption{Classification of IEs in the two immunology trials}
\label{tab::supp_ie_classification} \\
\toprule
IE category & Count & Proportion & Informative? & Rationale \\
\midrule
\endfirsthead

\multicolumn{5}{l}{\textit{Table \thetable{} continued}} \\
\toprule
IE category & Count & Proportion & Informative? & Rationale \\
\midrule
\endhead

\bottomrule
\endfoot

Adverse event & 135 & 31.5\% & Y & Classified as effect-informative because adverse events are plausibly related to treatment tolerability and the patient's underlying treatment response. \\
\hline

Others & 84 & 19.6\% & Y & For discontinuations with insufficiently detailed documentation, we classified the event as effect-informative because the absence of clear information makes an effect-uninformative interpretation difficult to justify. We adopted this default to avoid overstating arm-specific mean outcomes. \\
\hline

Lack of efficacy & 64 & 14.9\% & Y & Classified as effect-informative because lack of efficacy is directly informative about the patient's response to treatment. \\
\hline

Personal or family reason & 32 & 7.5\% & N & Review of the detailed site-level comments suggested that these discontinuations were driven by external personal circumstances rather than health status or treatment response. Examples included relocation of residence and other family-related reasons unrelated to disease progression or adverse events. \\
\hline

Protocol deviation & 26 & 6.1\% & Y & Classified as effect-informative because these deviations were typically linked to treatment management or patient status in ways that could reflect the patient's response trajectory. Examples included changes in corticosteroid use or prohibited dose changes in concomitant medication. \\
\hline

Due to epidemic or pandemic & 23 & 5.4\% & N & Classified as effect-uninformative because these discontinuations were driven by external disruptions, such as pandemic-related restrictions, that are more plausibly unrelated to treatment efficacy. \\
\hline

Withdrawal by subject & 22 & 5.1\% & Y & Classified as effect-informative because many such withdrawals were associated with concerns about side effects, unwillingness to continue treatment, or other reasons potentially related to treatment experience. When documentation was limited, we again defaulted to effect-informative classification. \\
\hline

Pregnancy & 15 & 3.5\% & N & Classified as effect-uninformative when the documentation suggested that treatment discontinuation was driven by pregnancy planning, suspected pregnancy, or pregnancy itself, rather than by the patient's disease course or treatment response. \\
\hline

Discontinued: unable to contact & 11 & 2.6\% & Y & Classified as effect-informative because the available documentation did not allow us to rule out discontinuation related to efficacy, adverse events, or other treatment-related concerns. \\
\hline

Death & 9 & 2.1\% & Y & Classified as effect-informative because death is directly informative about the patient's health status and precludes meaningful continuation of treatment. \\
\hline

Prohibited medication use & 5 & 1.2\% & Y & Classified as effect-informative because the need for prohibited medication may reflect inadequate disease control, toxicity management, or other clinically relevant aspects of treatment response. \\
\hline

Physician decision & 2 & 0.5\% & Y & Classified as effect-informative because physician-directed discontinuation was based on patient behavior or clinical status and therefore plausibly related to treatment experience or outcome risk. \\
\hline

Site terminated by sponsor & 1 & 0.2\% & N & Classified as effect-uninformative because sponsor-driven site closure is an external administrative disruption unrelated to the patient's biological treatment response. \\
\end{longtable}

\subsection{Exploratory data analysis: baseline covariates and IE summaries}
\label{sec::supp_eda}

\begin{figure}[h]
\centering
\includegraphics[width=\textwidth]{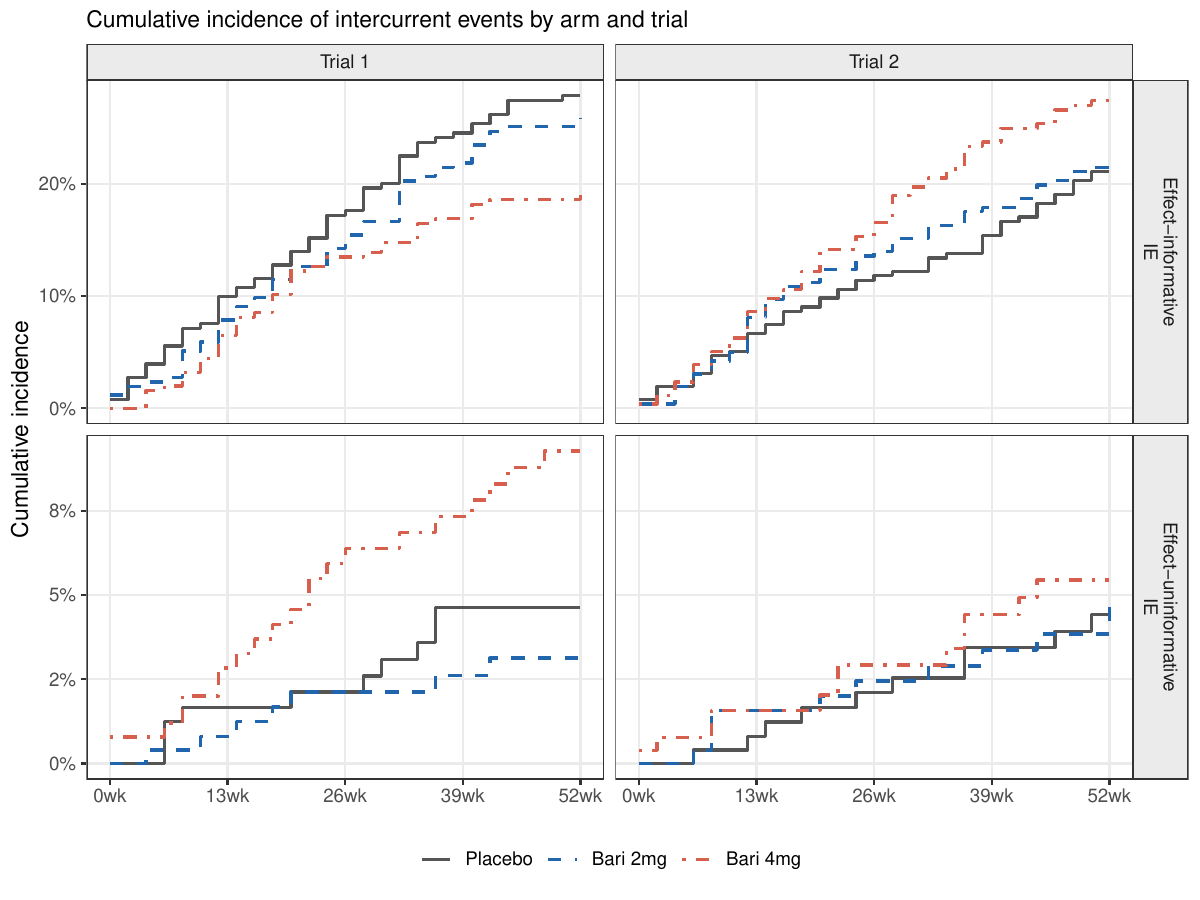}
\caption{Kaplan--Meier-based cumulative incidence of effect-informative IEs (top row) and effect-uninformative IEs (bottom row) by treatment arm, for Trial~1 (left column) and Trial~2 (right column). The $x$-axis shows weeks from treatment initiation and the $y$-axis shows cumulative incidence.}
\label{fig::supp_cumincidence}
\end{figure}

Table~\ref{tab::supp_baseline} reports summary statistics of baseline covariates and IEs for all six arm-by-trial combinations. Figure~\ref{fig::supp_cumincidence} shows the Kaplan--Meier-based cumulative incidence of effect-informative and effect-uninformative IEs by treatment arm and trial.

\begin{table}[h]
\caption{Baseline characteristics and IE summary by treatment arm and trial. PGA denotes Physician's Global Assessment score (scale 0--100); SLEDAI denotes the SLE Disease Activity Index. IE counts and percentages are relative to the full arm $n$. Median time to IE (in weeks) and interquartile range [Q1, Q3] are reported for participants who experienced the respective IE type.}
\doublespacing
\centering
\begin{small}
\begin{tabular}{lcccccc}
\\
\hline\hline
 & \multicolumn{3}{c}{Trial 1} & \multicolumn{3}{c}{Trial 2} \\
\cline{2-4}\cline{5-7}
 & Placebo & 2mg & 4mg & Placebo & 2mg & 4mg \\
\hline
$n$ & 253 & 255 & 252 & 256 & 261 & 258 \\
\multicolumn{7}{l}{\textit{Baseline covariates}} \\
PGA, mean (SD)       & 59.9 (15.0) & 60.5 (14.4) & 59.2 (15.8) & 60.0 (14.6) & 61.0 (12.5) & 58.7 (14.6) \\
SLEDAI, mean (SD) & 10.0 (3.2)  & 10.3 (3.4)  & 10.0 (2.9)  & 10.1 (3.2)  & 10.1 (3.4)  & 10.1 (3.0)  \\
\multicolumn{7}{l}{\textit{Region, $n$ (\%)}} \\
Asia                       & 31 (12.3) & 32 (12.5) & 31 (12.3) & 31 (12.1) & 33 (12.6) & 31 (12.0) \\
C/S America \& Mexico      & 74 (29.2) & 73 (28.6) & 72 (28.6) & 59 (23.0) & 58 (22.2) & 57 (22.1) \\
Europe                     & 64 (25.3) & 64 (25.1) & 63 (25.0) & 55 (21.5) & 57 (21.8) & 57 (22.1) \\
North America              & 47 (18.6) & 49 (19.2) & 50 (19.8) & 50 (19.5) & 50 (19.2) & 49 (19.0) \\
Rest of World              & 37 (14.6) & 37 (14.5) & 36 (14.3) & 61 (23.8) & 63 (24.1) & 64 (24.8) \\
\hline
\multicolumn{7}{l}{\textit{Intercurrent events}} \\
Effect-informative, $n$ (\%)          & 69 (27.3) & 65 (25.5) & 46 (18.3) & 53 (20.7) & 55 (21.1) & 70 (27.1) \\
\quad median (weeks) [Q1, Q3]             & 20 [8, 31]  & 23 [12, 32] & 17 [12, 29] & 21 [11, 40] & 16 [11, 32] & 20 [11, 33] \\
Effect-uninformative, $n$ (\%)        & 10 (4.0)  & 7 (2.7)   & 21 (8.3)  & 10 (3.9)  & 11 (4.2)  & 12 (4.7)  \\
\quad median (weeks) [Q1, Q3]             & 23 [7, 32]  & 18 [11, 28] & 20 [11, 31] & 25 [15, 36] & 24 [8, 41]  & 22 [8, 36]  \\
No IE, $n$ (\%)                       & 174 (68.8) & 183 (71.8) & 185 (73.4) & 193 (75.4) & 195 (74.7) & 176 (68.2) \\
\hline
\end{tabular}
\end{small}
\label{tab::supp_baseline}
\end{table}

\subsection{Robustness check: accelerated failure time model for nuisance estimation}
\label{sec::supp_aft}

Table~\ref{tab::supp_aft} replicates Table~\ref{tab::application_result} with all four Cox proportional hazards nuisance models replaced by log-normal accelerated failure time (AFT) models. All other components (propensity score logistic regression, outcome logistic regression, and martingale integral computation) are identical to the main analysis. Standard errors and confidence intervals are based on 500 bootstrap replicates.

\begin{table}[h]
\caption{Treatment effect of Baricitinib on the primary outcome SRI4 under the proposed hybrid estimand, using log-normal AFT models in place of Cox proportional hazards regression for all survival nuisance functions. The first four columns report results for Trial~1 and the last four columns for Trial~2. Within each trial, the first four rows report the point estimate, bootstrap standard error, $p$-value, and lower and upper 95\% confidence limits for each estimator of the 2mg treatment arm, and the last four rows report those of the 4mg treatment arm. For each trial and each treatment arm, we report the results based on all four estimators.}
\doublespacing
\centering
\begin{small}
\begin{tabular}{ccccccccccc}
\\
\hline\hline
 & & \multicolumn{4}{c}{Trial 1} && \multicolumn{4}{c}{Trial 2}\\
\cline{3-6}\cline{8-11}
 & & $\hat{\tau}^{\textup{out}}$ & $\hat{\tau}^{\textup{ipw}}$ & $\hat{\tau}^{\textup{aug}}$ & $\hat{\tau}^{\textup{eif}}$ && $\hat{\tau}^{\textup{out}}$ & $\hat{\tau}^{\textup{ipw}}$ & $\hat{\tau}^{\textup{aug}}$ & $\hat{\tau}^{\textup{eif}}$ \\
\hline
2mg & point    & 0.032 & 0.028 & 0.025 & 0.026 & & 0.017 & 0.019 & 0.023 & 0.022 \\
    & se       & 0.042 & 0.049 & 0.049 & 0.042 & & 0.042 & 0.069 & 0.069 & 0.042 \\
    & $p$-value & 0.449 & 0.571 & 0.602 & 0.537 & & 0.691 & 0.780 & 0.741 & 0.608 \\
    & CI lower & $-0.050$ & $-0.068$ & $-0.070$ & $-0.057$ & & $-0.066$ & $-0.116$ & $-0.112$ & $-0.061$ \\
    & CI upper & $0.114$ & $0.124$ & $0.121$ & $0.109$ & & $0.099$ & $0.155$ & $0.158$ & $0.104$ \\
\hline
4mg & point    & 0.108 & 0.119 & 0.114 & 0.113 & & $-0.005$ & $-0.004$ & $-0.004$ & $-0.002$ \\
    & se       & 0.046 & 0.046 & 0.046 & 0.046 & & 0.041 & 0.045 & 0.045 & 0.042 \\
    & $p$-value & 0.019 & 0.010 & 0.014 & 0.013 & & 0.911 & 0.933 & 0.938 & 0.969 \\
    & CI lower & $0.018$ & $0.029$ & $0.023$ & $0.023$ & & $-0.086$ & $-0.092$ & $-0.092$ & $-0.083$ \\
    & CI upper & $0.198$ & $0.209$ & $0.204$ & $0.203$ & & $0.076$ & $0.084$ & $0.085$ & $0.080$ \\
\hline
\end{tabular}
\end{small}
\label{tab::supp_aft}
\end{table}

\subsection{Robustness check: alternative IE classifications}
\label{sec::robustness_ie_class}

As a further robustness check on the IE classification, we tested three alternative classification rules for borderline discontinuation categories. The primary analysis treats ``Withdrawal by Subject'' and ``Discontinued: Unable to Contact'' as effect-informative, and ``Pregnancy'' as effect-uninformative. The three robustness check scenarios are: (S1)~both ``Withdrawal by Subject'' and ``Discontinued: Unable to Contact'' reclassified as effect-uninformative; (S2)~``Withdrawal by Subject'' reclassified as effect-uninformative; (S3)~``Pregnancy'' reclassified as effect-informative. Table~\ref{tab::robustness_ie_class} reports the treatment effect estimates for the primary SRI4 outcome under the three alternative IE classification rules. Each scenario was analyzed using Cox proportional hazards nuisance models, with 200 nonparametric bootstrap replicates. All four estimators are shown alongside the bootstrap standard error, $p$-value, and 95\% confidence interval, for the EIF estimator.

\begin{table}[h]
\caption{IE classification robustness check for the primary outcome SRI4. Rows within each scenario panel correspond to the four pairwise comparisons. Columns show all four point estimators $\hat\tau^{\textup{out}}$, $\hat\tau^{\textup{ipw}}$, $\hat\tau^{\textup{aug}}$), and $\hat\tau^{\textup{eif}}$, together with the bootstrap standard error, 95\% confidence interval, and $p$-value for $\hat\tau^{\textup{eif}}$.}
\doublespacing
\centering
\begin{small}
\begin{tabular}{ccccccccc}
\\
\hline\hline
 & Treatment arm & $\hat\tau^{\textup{out}}$ & $\hat\tau^{\textup{ipw}}$ & $\hat\tau^{\textup{aug}}$ & $\hat\tau^{\textup{eif}}$ & se & 95\% CI & $p$-value \\
\hline
\multirow{4}{*}{S1}
 & Trial 1, 2mg & 0.031 & 0.028 & 0.026 & 0.026 & 0.045 & $[-0.062,\,0.113]$ & 0.565 \\
 & Trial 1, 4mg & 0.107 & 0.112 & 0.107 & 0.105 & 0.047 & $[0.012,\,0.198]$ & 0.027 \\
 & Trial 2, 2mg & 0.015 & 0.013 & 0.017 & 0.016 & 0.045 & $[-0.071,\,0.104]$ & 0.712 \\
 & Trial 2, 4mg & 0.004 & 0.004 & 0.004 & 0.005 & 0.043 & $[-0.079,\,0.089]$ & 0.908 \\
\hline
\multirow{4}{*}{S2}
 & Trial 1, 2mg & 0.028 & 0.026 & 0.024 & 0.023 & 0.044 & $[-0.064,\,0.109]$ & 0.608 \\
 & Trial 1, 4mg & 0.103 & 0.109 & 0.104 & 0.101 & 0.047 & $[0.009,\,0.194]$ & 0.031 \\
 & Trial 2, 2mg & 0.018 & 0.017 & 0.021 & 0.020 & 0.045 & $[-0.068,\,0.108]$ & 0.656 \\
 & Trial 2, 4mg & 0.003 & 0.004 & 0.004 & 0.005 & 0.043 & $[-0.080,\,0.090]$ & 0.911 \\
\hline
\multirow{4}{*}{S3}
 & Trial 1, 2mg & 0.031 & 0.030 & 0.028 & 0.028 & 0.044 & $[-0.059,\,0.115]$ & 0.525 \\
 & Trial 1, 4mg& 0.105 & 0.111 & 0.106 & 0.105 & 0.046 & $[0.014,\,0.195]$ & 0.023 \\
 & Trial 2, 2mg & 0.013 & 0.012 & 0.016 & 0.015 & 0.044 & $[-0.071,\,0.102]$ & 0.727 \\
 & Trial 2, 4mg & $-0.006$ & $-0.005$ & $-0.005$ & $-0.005$ & 0.044 & $[-0.090,\,0.081]$ & 0.910 \\
\hline
\end{tabular}
\end{small}
\label{tab::robustness_ie_class}
\end{table}

\subsection{Results on causal risk ratio and odds ratio}
\label{sec::supp_rr_or}

Table~\ref{tab::supp_rr_or} reports the causal risk ratio (\textsc{rr}) and odds ratio (\textsc{or}) for the primary SRI4 outcome under the proposed hybrid estimand. Both are estimated using the EIF-based estimator of the arm-specific mean composite outcomes $\hat{\mu}_1$ and $\hat{\mu}_0$. Confidence intervals for \textsc{rr} and \textsc{or} are constructed on the log scale using the normal approximation and then exponentiated.

\begin{table}[h]
\caption{Causal risk ratio and odds ratio for the primary outcome SRI4 under the proposed hybrid estimand. $\hat\tau$, $\hat{\textsc{rr}}$, and $\hat{\textsc{or}}$ denote the estimated original estimand $\tau$ (risk difference), risk ratio, and odds ratio, respectively. Confidence intervals for \textsc{rr} and \textsc{or} are constructed on the log scale and exponentiated.}
\doublespacing
\centering
\begin{small}
\begin{tabular}{ccccc}
\\
\hline\hline
Treatment arm & $\hat\tau$ (95\,\% CI) & $\hat{\textsc{rr}}$ (95\,\% CI) & $\hat{\textsc{or}}$ (95\,\% CI) \\
\hline
Trial 1, 2mg & $0.026\;[-0.057,\,0.110]$    & $1.06\;[0.88,\,1.27]$     & $1.11\;[0.79,\,1.56]$     \\
Trial 1, 4mg & $0.113\;[0.023,\,0.203]$ & $1.25\;[1.04,\,1.49]$ & $1.58\;[1.09,\,2.27]$ \\
Trial 2, 2mg & $0.022\;[-0.061,\,0.104]$    & $1.05\;[0.88,\,1.25]$     & $1.09\;[0.78,\,1.52]$     \\
Trial 2, 4mg & $-0.002\;[-0.084,\,0.080]$   & $1.00\;[0.83,\,1.19]$     & $0.99\;[0.71,\,1.39]$     \\
\hline
\end{tabular}
\end{small}
\label{tab::supp_rr_or}
\end{table}

\subsection{Computational time}
\label{sec::supp_timing}

Table~\ref{tab::supp_timing} reports the time for a single full estimator evaluation on each of the four pairwise comparison datasets, measured on a MacBook Air with Apple M4 processor. The full procedure includes propensity score logistic regression, four Cox proportional hazards survival models (two for effect-informative IE, two for effect-uninformative IE), six outcome logistic regression models (one per outcome per treatment arm), and the discrete-time martingale integral required by the EIF-based estimator. The estimated bootstrap time is the single-run time times 500 replicates.

\begin{table}[h]
\caption{Wall-clock time for a single full-estimator evaluation and the estimated time for 500 nonparametric bootstrap replicates. Sample size $n$ includes both the treatment and placebo arms for each pairwise comparison.}
\doublespacing
\centering
\begin{small}
\begin{tabular}{cccc}
\\
\hline\hline
Treatment arm & $n$ & Single run (sec) & Bootstrap (min) \\
\hline
Trial 1, 2mg & 508 & 0.32 & 2.7 \\
Trial 1, 4mg & 505 & 0.31 & 2.6 \\
Trial 2, 2mg & 517 & 0.41 & 3.4 \\
Trial 2, 4mg & 514 & 0.34 & 2.8 \\
\hline
\end{tabular}
\end{small}
\label{tab::supp_timing}
\end{table}

\section{Simulation}
\label{sec::simulation}
In this section, we conduct Monte Carlo simulations to study the finite sample performance of our proposed estimators. 
\subsection{Data generating processes}
We first generate the covariates $X=(X_1,X_2,X_3)^{\T}\in \mathbb{R}^3$ from three independent standard Gaussian distributions and denote $\tilde{X}_j=\{(X_j+2)^2-1\}/\sqrt{12}$ for $j=1,2,3$. Next, we generate the treatment assignment following $A\mid X\sim \Bern(\{e(X)\})$ with $e(X)$ being the propensity score model, and generate the potential outcomes following $Y(a,c=\infty)\mid X\sim \mathcal{N}(\mu_a(X),\sigma_a^2)$ with $\mu_a(X)$ being the outcome model for $a=0,1$. We then generate the potential values of the effect-informative IE survival time $T(a)$ from the distribution with a survival function $$S_a(t\mid X)=\exp[-0.002t^{1.2}\exp\{\gamma_a(X)\}]$$ and the potential values of the effect-uninformative IE survival time $C(a)$ from the distributions with a survival function $$G_a(t\mid X)=\exp[\varrho_a(t)\exp\{\delta_a(X)\}],$$ for $a=0,1$. For each of the nuisance parameters, we consider two scenarios when it will be correctly modeled and misspecified, therefore, we generate data following two different choices of each. We summarize the detailed data-generating choices in Table~\ref{tab::dgp_nuisance}.

\begin{table}[h]
\doublespacing
\centering
\caption{Model choices for the nuisance parameters}
\begin{small}
\begin{tabular}{lll}
\\
\hline
\hline
&  Correctly specified & Misspecified\\
\hline
 $\textup{logit}\{e(X)\}$& $(X_1+X_2+X_3)/5$ & $1(X_1\geq 0)\{\exp(\tilde X_2) - X_2(1+\tilde X_3)\}-\exp(\tilde X_2)$ \\
 $\mu_1(X)$& $2(X_1+X_2+X_3)$ & $1(X_1\geq 0)\{X_2 + \exp(X_2)\tilde X_3 - \tilde X_2\} + \tilde X_2$\\
 $\mu_0(X)$& $X_1+X_2+X_3$ & $-\tilde X_1 - 1(X_1>0.5)\tilde X_2 + 1(X_1<-0.5)X_2^2\log(|X_3|+1)$\\
 $\sigma_a$& $0.1(a+1)$& $1$\\
 $\gamma_1(X)$&  $0.1(X_1+2X_2-2X_3)$& $0.1(X_1^2X_2-X_2-1)+X_2\log(10X_3^2)$\\
 $\gamma_0(X)$&  $0.1(X_1-2X_2+2X_3)$& $0.01(-\tilde X_1+\tilde X_2+\tilde X_3)$\\
 $\delta_1(X)$&  $0$& $0.1(X_1^2X_2-X_2-1)+X_2\log(10X_3^2)$\\
 $\delta_0(X)$&  $0$& $0$\\
 $\varrho_a(t)$& $-0.01t^{1.2}$& $-0.01at^{1.2}+0.6*0.01^{1/1.2}(1-a)t$ \\
\hline
\end{tabular}
\end{small}
\label{tab::dgp_nuisance}
\end{table}

We consider five data-generating regimes: all four models are correctly specified, misspecified $e$ and correct $(G,\mu,S)$, misspecified $(e, G)$ and correct $(\mu, S)$, misspecified $(\mu, S)$ and correct $(e,G)$, and all four models are misspecified. Let $Y=AY(1,c=\infty)+(1-A)Y(0,c=\infty)$, $T=AT(1)+(1-A)T(0)$, and $C=AC(1)+(1-A)C(0)$. Generate the observed event time as $T\wedge C\wedge k$, the observed event type indicator, and the observed outcome $Y$ if $T\wedge C>k$.

\subsection{Simulation results}
We compare the finite sample performance of the outcome estimator $\hat\tau^{\textup{out}}$, the weighting estimator $\hat\tau^{\textup{ipw}}$, the augmented weighting estimator $\hat\tau^{\textup{aug}}$, and the EIF estimator $\hat\tau^{\textup{eif}}$ based on $1000$ Monte Carlo samples with a sample size of $n=1000$. When fitting the nuisance models, we use logistic regression of $A$ on $X$ to estimate the propensity score model, the linear regression of $Y$ on $X$ on the subsample $A=a, T\wedge C>k$ to estimate the outcome model, and the Cox proportional hazard regression of $T$ and $C$ on $X$ on the subsample $A=a$ to estimate the effect-informative IE survival model and the censoring model, respectively, for $a=0,1$. We use $X$ in all regression models, therefore, when the true data-generating processes involve non-linear functions of $X$, the fitted nuisance models suffer from misspecification. We evaluate the performance of the four estimators with reported bias, standard deviation, and coverage probability using a nonparametric bootstrap with $200$ bootstrap samples in Table~\ref{tab::simu_res}.

\begin{table}[h]
\caption{Finite sample performance of the four estimators. For each estimator, we report the finite-sample bias, standard deviation (SD), and the coverage rate (CR) of a 95\% confidence interval, which is constructed using a nonparametric bootstrap with 200 bootstrap iterations. Each row corresponds to a different data-generating regime.}
\doublespacing
\centering
\begin{tiny}
\begin{tabular}{lrcclrcclrcclrcc}
\\
\hline
\hline
& \multicolumn{3}{c}{$\hat\tau^{\textup{out}}$}& &\multicolumn{3}{c}{$\hat\tau^{\textup{ipw}}$}& & \multicolumn{3}{c}{$\hat\tau^{\textup{aug}}$} & &\multicolumn{3}{c}{$\hat\tau^{\textup{eif}}$}\\
\cline{2-4} \cline{6-8} \cline{10-12} \cline{14-16}
& Bias&  SD&  CR&  &   Bias& SD& CR& & Bias&SD&  CR&  &  Bias& SD&CR\\
\hline
    all\_correct & 0.003& 0.100& 0.989& & 0.005& 0.154& 0.977& & 0.004 & 0.152 & 0.980&  &  0.004& 0.110&0.982\\
    $e$\_wrong & $-0.002$&  0.134& 0.978& & $-0.338$& 0.527& 0.916& & $-0.003$&0.479&  0.961&  &  0.000& 0.234&0.979\\
    $e\_G$\_wrong & $-0.002$& 0.130& 0.985& & $-0.194$& 1.591& 0.831& & 0.141& 1.559& 0.972& & $-0.013$& 0.354&0.969\\
    $\mu\_S$\_wrong &$-0.085$& 0.170& 0.936& & $-0.009$& 0.186& 0.974& & $-0.009$& 0.186& 0.974& & $-0.012$& 0.176&0.972\\
    all\_wrong & 0.230& 0.233& 0.837& & 0.660& 5.799& 0.898& & 0.850& 5.782& 0.967& & 0.573& 3.022&0.919\\
\hline
\end{tabular}
\end{tiny}
\label{tab::simu_res}
\end{table}

Consistent with our theoretical results, all estimators have a finite sample bias close to zero when all models are correctly specified. When $e$ is misspecified and $(G,\mu,S)$ are correct, the weighting estimator $\hat\tau^{\textup{ipw}}$ is inconsistent while the other three estimators have small finite sample bias. Under the regime when $e$ and $G$ are misspecified, both $\hat\tau^{\textup{ipw}}$ and $\hat\tau^{\textup{aug}}$ perform poorly as they have larger finite sample biases, $\hat\tau^{\textup{out}}$ and $\hat\tau^{\textup{eif}}$ show small finite sample biases as expected. When $(\mu,S)$ are misspecified, the outcome estimator $\hat\tau^{\textup{out}}$ has a relatively large bias, while all other three estimators have near-zero biases. All estimators have non-negligible finite sample bias when all nuisance models are misspecified. $\hat\tau^{\textup{eif}}$ is most robust to model misspecifications across all different regimes. 

$\hat\tau^{\textup{eif}}$ shows a relatively smaller standard deviation compared to $\hat\tau^{\textup{ipw}}$ and $\hat\tau^{\textup{aug}}$ in the regimes where all estimators are consistent. The outcome estimator $\hat\tau^{\textup{out}}$ always has the smallest standard deviation when $(\mu,S)$ is correctly specified. Both $\hat\tau^{\textup{ipw}}$ and $\hat\tau^{\textup{aug}}$ have larger standard deviations across all regimes in which they are consistent. Furthermore, confidence intervals constructed using the nonparametric bootstrap yield valid coverage rates for the corresponding consistent estimators in their respective regimes.

Next, we evaluate two ad-hoc methods commonly used in clinical trials with IEs, demonstrating through simulations that they are biased in estimating the clinically relevant causal parameter. The first method is the non-responder rule method that assigns an outcome as $0$ whenever an IE occurs. It is equivalent to using the composite outcome strategy that treats all IEs as effect-informative, leading to a composite outcome equal to $0$. The second method naively applies a hypothetical strategy to all IEs, assuming they are independent of potential outcomes conditional on observed covariates. We generate data under the ``$\mu\_S$\_wrong'' regime, ensuring correct specification of both the propensity score $e(X)$ and the effect-uninformative IE survival model $G_a(t\mid X)$ for $a=0,1$. Based on $1000$ Monte Carlo simulations with a sample size of $n=1000$, the inverse probability-weighting estimator using the non-responder rule method exhibits a finite sample bias of $-1.195$ with a standard deviation of $0.077$, and that using the hypothetical strategy for all IEs has a bias of $1.188$ with a standard deviation of $0.388$. Both ad-hoc methods yield inconsistent estimators. 

\section{Proofs}
\label{sec::proofs}
\subsection{Proof of Theorem~\ref{thm::comp_out_identification}}
First, equation~\eqref{eqn::id_out} holds because its left-hand side
\begin{eqnarray*}
    && E\{Y(1,c=\infty)1(T(1) > k)\} \\
    &=& E[E\{Y(1,c=\infty)1(T(1) > k)\mid X\}] \\
    &=& E[E\{Y(1,c=\infty)1(T > k)\mid X,A=1\}] \\
    &=& E\{E(Y(1,c=\infty)\mid T>k,X,A=1) \pr(T>k\mid X,A=1)\} \\
    &=& E\{E(Y(1,c=\infty)\mid C>k,T>k,X,A=1) \pr(T>k\mid X,A=1)\} \\
    &=& E\{E(Y\mid C>k,T>k,X,A=1) \pr(T>k\mid X,A=1)\},
\end{eqnarray*}
which is equal to the right-hand side of~\eqref{eqn::id_out}, where the first equality is by the law of iterated expectations, the second equality is by the randomization assumption~\ref{assump::randomization}, and the fourth equality is by the censoring at random assumption~\ref{assump::car}.

Next, equation~\eqref{eqn::id_ipw} holds because its right-hand side equals
\begin{eqnarray*}
    &&E\left[ E\left\{ \frac{AY1(T\wedge C>k)}{e(X)\pr(C>k\mid X, A=1)} \mid X \right\} \right] \\
    &=& E\left( \frac{1}{e(X)\pr(C>k\mid X, A=1)} E\left[ AY(1,c=\infty)1\{T(1)\wedge C(1)>k\} \mid X \right] \right) \\
    &=& E\left( \frac{1}{e(X)\pr(C>k\mid X, A=1)} E(A\mid X) E\left[Y(1,c=\infty)1\{T(1)>k\} 1\{C(1)>k\} \mid X \right] \right)\\
    &=& E\left( \frac{1}{\pr(C>k\mid X, A=1)} E[1\{C(1)>k\}\mid X] E\left[Y(1,c=\infty)1\{T(1)>k\} \mid X \right] \right) \\
    &=& E[E\{Y(1,c=\infty)1(T(1) > k)\mid X\}],
\end{eqnarray*}
which is equal to the left-hand side of~\eqref{eqn::id_ipw}, where the first equality is by the law of iterated expectations, the third equality is by the randomization assumption~\ref{assump::randomization}, the fourth equality is by the censoring at random assumption~\ref{assump::car}, and the last equality is again by the law of iterated expectations.
\QEDB

\subsection{Proof of Proposition~\ref{prop::tau_aug}}
We prove that under the correct specification of the censoring model, $\hat\tau^{\textup{aug}}$ is doubly robust in the sense that it is consistent for $\tau$ if either $e^*=e$ or $(\mu_1^* = \mu_1, S_1^*=S_1)$. We have
\begin{eqnarray}
    && E\left\{\frac{AY1(T\wedge C>k)}{e^*(X)G_1(k\mid X)} - \frac{A-e^*(X)}{e^*(X)}\mu_1^*(X)S_1^*(k\mid X)\right\} \notag \\
    &=& E\left[\frac{E\{AY1(T\wedge C>k)\mid X\}}{e^*(X)G_1(k\mid X)} - \frac{e(X)-e^*(X)}{e^*(X)}\mu_1^*(X)S_1^*(k\mid X)\right] \notag \\
    &=& E\left\{\frac{e(X)}{e^*(X)}\mu_1(X)S_1(k\mid X) - \frac{e(X)-e^*(X)}{e^*(X)}\mu_1^*(X)S_1^*(k\mid X)\right\}, \label{eqn::proof_prop_tau_aug}
\end{eqnarray}
where the first equality follows from the law of integrated expectations and the second equality follows from similar derivations as in the proof of Theorem~\ref{thm::comp_out_identification}. Now observe that the final expression of~\eqref{eqn::proof_prop_tau_aug} is equal to $\mu_1$ if either $e^*=e$ or $(\mu_1^* = \mu_1, S_1^*=S_1)$. Therefore, we have
\begin{eqnarray*}
    E\left\{\frac{AY1(T\wedge C>k)}{e^*(X)G_1(k\mid X)} - \frac{A-e^*(X)}{e^*(X)}\mu_1^*(X)S_1^*(k\mid X) - \mu_1\right\} &=& 0
\end{eqnarray*}
if either $e^* = e$ or $(\mu_1^* = \mu_1, S_1^*=S_1)$ holds. A parallel argument applies to the control counterpart. Therefore, combining the treated and control components, we conclude that $\hat\tau^{\textup{aug}}$ is consistent for $\tau$ if either the propensity score model is correct, or the outcome and survival models for the effect-informative IE are both correct. This establishes the double robustness of $\hat\tau^{\textup{aug}}$.
\QEDB

\subsection{Proof of Theorem~\ref{thm::eif_tau}}
\label{sec::proof_eif_tau}
We follow the semiparametric theory in \cite{bickel1993efficient} to derive the EIF for $\mu_1$. As discussed in Section~\ref{sec::eif}, we have two levels of coarsening of the full data $(X, Y^\c(1), Y^\c(0))$, one is because the effect-uninformative IE time $C(a)$ is censoring $T(a)\wedge k$, and the other is due to the treatment assignment. Following the steps in \cite{hubbard2000nonparametric}, we first consider the case when every observation is assigned to the treatment group, i.e., when there is no missingness generated by the treatment assignment, and then project the derived EIF onto the nuisance tangent space of the propensity score to get the final form of EIF for $\mu_1$.

\paragraph{Step 1.}
With full data $(X, Y^\c(1), Y^\c(0))$, the estimating equation for $\mu_1$ is $D_{\textup{full}}=Y^\c(1) - \mu_1$. Since $Y^\c(1)$ is only observable if $C(1)>T(1)\wedge k$, the full data is not available even if all corresponding potential outcomes under treatment $A=1$ are observed, and the observed data is 
\begin{eqnarray*}
    O(1) &=& (X, \Delta(1), \tilde{T}(1), \Delta(1)Y^\c(1)) \sim P_1,
\end{eqnarray*} 
where $\Delta(1)=1(C(1)>T(1)\wedge k)$ is the missing indicator with $\Delta(1)=1$ if $Y^\c(1)$ is observed and $\Delta(1)=0$ otherwise, and $\tilde{T}(1)=T(1)\wedge C(1)\wedge k$ is the observed event time with $\tilde{T}(1)=T(1)\wedge k$ if $\Delta(1)=1$, $Y^\c(1)$ is observed, and $\tilde{T}(1)=C(1)$ if $\Delta(1)=0$, $Y^\c(1)$ is missing. An identification formula for $\mu_1$ is 
\begin{equation*}
    \mu_1 = E\left\{ \frac{\Delta(1) Y^\c(1)}{G_1(\tilde{T}(1)\mid X)} \right\},
\end{equation*}
where $G_1(t\mid X)=\pr(C(1)>t\mid X)$ is the probability of not censoring up to time $t$ conditioning on the covariates. We can write the IPCW estimating equation as
\begin{equation*}
    D_{\textup{IPCW}} = \frac{\Delta(1)\{Y^\c(1) - \mu_1\}}{G_1(\tilde{T}(1)\mid X)},
\end{equation*}
and have $E(D_{\textup{IPCW}}) = 0$ by the law of iterated expectations and the censoring at random assumption. 

Next, we follow the steps in \cite{van2007note} to derive the EIF for $\mu_1$ when the observed data is $O(1)=(X, \Delta(1), \tilde{T}(1), \Delta(1)Y^\c(1))$. For ease of notation, we omit the subscripts, superscripts, or numbers in parenthesis $(1)$ that indicate the potential outcomes under the treatment $A=1$ in the following derivation, so the dependence on the treatment arm is implicit. For ease of notation, denote $T^1=T\wedge k$.

Let $T(P)$ denote the tangent space which is the whole Hilbert space $L_0^2(P)$ since our model is nonparametric. The tangent space can be written as a direct sum of three components $T = T_X\oplus T_{\textup{F}}\oplus T_{\textup{CAR}}$ with
\begin{eqnarray*}
    T_X &=& \{h(X)\in L_0^2(P): E\{h(X)\}=0\},\\
    T_{\textup{F}} &=& \{h(O)\in L_0^2(P): E\{h(O) \mid C, X\}=0\},\\
    T_{\textup{CAR}} &=& \{h(O)\in L_0^2(P): E\{h(O) \mid Y^\c, T^1, X\}=0\},
\end{eqnarray*}
where $T_X$, $T_{\textup{F}}$, and $T_{\textup{CAR}}$ are orthogonal to each other due to the censoring at random assumption and the factorization of the observed data. These tangent spaces are generated from scores of submodels that perturb the marginal distribution of $X$, the conditional distribution of $Y^1\mid X$, and the conditional censoring probability $C\mid X$, respectively. Due to the orthogonality, we have the decomposition of any $h(O)\in L_0^2(P)$ as
\begin{eqnarray*}
    h(O) &=& \Pi \{ h \mid T(P) \} \\
    &=& \Pi \{ h \mid T_X(P) \} + \Pi \{ h \mid T_{\textup{F}}(P) \} + \Pi \{ h \mid T_{\textup{CAR}}(P) \}.
\end{eqnarray*}
Let $D(O)$ denote the efficient influence function for $\mu_1$. By the results in Chapter 1.4 of \cite{van2003unified}, we have: (1) $D(O)$ should be orthogonal to the nuisance tangent space $T_{\textup{CAR}}$, thus $\Pi \{ D(O) \mid T_{\textup{CAR}}(P)\}=0$; (2) $D(O)$ can be rewritten as
\begin{equation*}
D = D_{\textup{IPCW}} - \Pi\{ D_{\textup{IPCW}} \mid T_{\textup{CAR}}(P) \},
\end{equation*}
thus
\begin{eqnarray*}
    \Pi \{ D \mid T_{\textup{F}}(P) \} &=& \Pi \{ D_{\textup{IPCW}} \mid T_{\textup{F}}(P) \} - \Pi\{ \Pi\{ D_{\textup{IPCW}} \mid T_{\textup{CAR}}(P) \} \mid T_{\textup{F}}(P)\} \\
    &=& \Pi \{ D_{\textup{IPCW}} \mid T_{\textup{F}}(P) \}
\end{eqnarray*}
where the last equality is by the fact that $T_{\textup{F}}$ and $T_{\textup{CAR}}$ are orthogonal to each other, and similarly, $\Pi \{ D \mid T_X(P) \} = \Pi \{ D_{\textup{IPCW}} \mid T_X(P) \}$. 

These projections provide us with two different ways to compute the EIF $D$: (1) directly compute the projections $\Pi \{ D_{\textup{IPCW}} \mid T_X(P) \}$ and $\Pi \{ D_{\textup{IPCW}} \mid T_{\textup{F}}(P) \}$ and sum them up; (2) compute the projection $\Pi \{ D_{\textup{IPCW}} \mid T_{\textup{CAR}}(P) \}$ and subtract it from $D_{\textup{IPCW}}$. In the classic survival outcome problem, these two methods are symmetric since $T$ and $C$ are censoring each other thus the projections on $T_{\textup{F}}(P)$ and $T_{\textup{CAR}}(P)$ are very similar. However, due to the complication generated by $Y$ in our setting, the second approach is easier, since the tangent space $T_{\textup{F}}(P)$ is hard to compute.

We next compute $\Pi\{ D_{\textup{IPCW}} \mid T_{\textup{CAR}}(P) \}$. By Theorem 1.1 in \cite{van2003unified}, 
\begin{equation*}
   T_{\textup{CAR}}(P) = \overline{\left\{\int H(t,\mathcal{F}(t))\d M_G(t)\textup{ for all functions }H(t,\mathcal{F}(t))\right\} \cap L_0^2(P)},
\end{equation*}
and the projection of a function $h(O)$ onto $T_{\textup{CAR}}(P)$ is
\begin{equation*}
    \Pi\{ h(O) \mid T_{\textup{CAR}}(P) \} = \int_0^{\tilde{T}} \left\{ E(h(O)\mid \d A(t)=1, \mathcal{F}(t)) - E(h(O)\mid \d A(t)=0, \mathcal{F}(t)) \right\} \d M_G(t),
\end{equation*}
where $A(t)=1(C\leq t)$ is the indicator of whether censoring happens up until time $t$ (define $C=\infty$ if $C> T^1$ so that $C$ is always observed), $\mathcal{F}(t) = (\bar{A}(t-),X)$ is the history observed up to time $t$, and $\d M_G(t) = 1(C\in \d t, \Delta = 0) - 1(\tilde{T}\geq t)\d\Lambda(t\mid X)$ is the Doob--Meyer martingale of the counting process of censoring $C$. $E(D_{\textup{IPCW}}\mid \d A(t)=1, \mathcal{F}(t)) = 0$ by the definition of $D_{\textup{IPCW}}$, thus we only need to compute $E(D_{\textup{IPCW}}\mid \d A(t)=0, \mathcal{F}(t))$, which plus $\mu_1$ is equal to
\begin{eqnarray*}
    && E\left\{ \frac{\Delta Y^\c}{G(\tilde{T}\mid X)} \mid \d A(t)=0, \mathcal{F}(t) \right\}\\
    &=& E\left\{ \frac{\Delta Y^\c}{G(\tilde{T}\mid X)} \mid T^1\geq t, C\geq t, X \right\} \\
    &=& E\left\{ \frac{\Delta Y^\c}{G(\tilde{T}\mid X)} \mid T^1\geq t, X \right\} \big/ G(t\mid X) \\
    &=& E\left\{ \frac{Y^\c}{G(T\wedge k\mid X)} \pr(C>T\wedge k\mid T^1\geq t, T, Y^1, X) \mid T^1\geq t, X \right\} \big/ G(t\mid X) \\
    &=& E\left\{ \frac{Y^\c}{G(T\wedge k\mid X)} \pr(C>T\wedge k\mid X) \mid T^1\geq t, X \right\} \big/ G(t\mid X) \\
    &=& E\left\{ Y^\c \mid T^1\geq t, X \right\} \big/ G(t\mid X).
\end{eqnarray*}
By the fact that the integration is over $t:t<\tilde{T}$, we have $t<T\wedge k\wedge C$ thus $t>k$. The conditional expectation
\begin{eqnarray*}
    E\left\{ Y^\c\mid T^1\geq t, X \right\} &=& E\left\{ Y^\c\mid T^1\geq t, T^1=T, X \right\} \pr(T^1=T\mid T^1\geq t, X) \\
    && + E\left\{ Y^\c\mid T^1\geq t, T^1=k, X \right\} \pr(T^1=k\mid T^1\geq t, X)\\
    &=& E(Y\mid T>k\geq t, X) \pr(T>k\mid T^1\geq t, X) \\
    &=& E(Y\mid T>k, X) \pr(T>k\mid T\wedge k \geq t, X) \\
    &=& E(Y\mid T>k, X) \pr(T>k\mid T \geq t, X) \\
    &=& \frac{E(Y1(T>k)\mid X)}{\pr(T>k\mid X)} \frac{\pr(T>k\mid X)}{\pr(T\geq t \mid X)}\\
    &=& \frac{E(Y1(T>k)\mid X)}{\pr(T\geq t \mid X)}.
\end{eqnarray*}
Therefore, we have
\begin{eqnarray*}
    \Pi\{ D_{\textup{IPCW}} \mid T_{\textup{CAR}}(P) \} + \mu_1 &=& - \int_0^{\tilde{T}}  \frac{E(Y1(T>k)\mid X)}{\pr(T\geq t \mid X)} \frac{\d M_G(t)}{G(t\mid X)}\\
    &=& -E(Y1(T>k)\mid X)\int_0^{\tilde{T}} \frac{\d M_G(t)}{S(t\mid X) G(t\mid X)},
\end{eqnarray*}
and thus the EIF assuming $O(1)$ is the observed data is
\begin{eqnarray}
    D &=& \frac{\Delta Y^\c}{G(\tilde{T}\mid X)} + E(Y^\c\mid X)\int_0^{\tilde{T}}\frac{\d M_G(t)}{S(t\mid X) G(t\mid X)} - \mu_1 \label{eqn::eif_potential_outcome_1}
\end{eqnarray}

\paragraph{Step 2.} Next, we follow steps in Section 3 of \cite{hubbard2000nonparametric} and compute the EIF when the real observed data is $O=(X, A, \Delta, \tilde{T}, \Delta Y^\c)$. We need to construct a weighting estimating equation and then subtract its projection onto the nuisance tangent space of the propensity score to get the final form of the EIF. To be clear on the distinction between potential outcomes and the observed values, we add back the dependence on the treatment assignment in~\eqref{eqn::eif_potential_outcome_1} and write it as
\begin{eqnarray*}
    D(1) &=& \frac{\Delta(1) Y^\c(1)}{G_1(\tilde{T}(1)\mid X)} + E(Y^\c(1)\mid X)\int_0^{\tilde{T}(1)}\frac{\d M_{G_1}(t)}{S_1(t\mid X) G_1(t\mid X)} - \mu_1.
\end{eqnarray*}
A valid weighting estimating equation is
\begin{eqnarray*}
    D_{\textup{IPW}} &=& \frac{A}{e(X)}\left\{\frac{\Delta Y^\c}{G_1(\tilde{T}\mid X)} + E(Y^\c\mid X, A=1)\int_0^{\tilde{T}}\frac{\d M_{G_1}(t)}{S_1(t\mid X) G_1(t\mid X)}\right\} - \mu_1.
\end{eqnarray*}
Further, project this onto the nuisance tangent space of the propensity score, the projection is
\begin{eqnarray*}
    \Pi\{D_{\textup{IPW}}\mid T_{\textup{pscore}}\} &=& E\left\{D_{\textup{IPW}}\mid A, X\right\} - E\left\{D_{\textup{IPW}}\mid X\right\} \\
    &=& \frac{A-e(X)}{e(X)}E\left\{\frac{\Delta Y^\c}{G_1(\tilde{T}\mid X)}\mid X, A=1\right\} \\
    &=& \frac{A-e(X)}{e(X)}E(Y^\c\mid X, A=1),
\end{eqnarray*}
where the second equality is by the fact that $E\{\d M_{G_1}(t)\mid X, A=1\}=E\{\d M_{G_1}(t)\mid X\}=0$, and the last equality is by the censoring at random assumption. Therefore, the EIF 
\begin{eqnarray*}
    D_{1} &=& D_{\textup{IPW}} - \Pi\{D_{\textup{IPW}}\mid T_{\textup{pscore}}\} \\
    &=& \frac{A}{e(X)}\left\{\frac{\Delta Y^\c}{G_1(\tilde{T}\mid X)} + E(Y1(T>k)\mid X, A=1)\int_0^{\tilde{T}}\frac{\d M_{G_1}(t)}{S_1(t\mid X) G_1(t\mid X)}\right\} \\
    && - \frac{A-e(X)}{e(X)}E(Y1(T>k)\mid X, A=1) - \mu_1
\end{eqnarray*}
has the given form in Theorem \ref{thm::eif_tau}.
\QEDB

\subsection{Proof of Theorem~\ref{thm::double_robustness}}
We prove the result by showing 
\begin{eqnarray}
    \mu_1 &=& E\left[\frac{A}{e^{*}(X)}\left\{\frac{Y1(T\wedge C>k)}{G_1^{*}(k\mid X)} + \mu_1^{*}(X)S_1^{*}(k\mid X)\int_{0}^{\tilde{T}}\frac{\d M_{G_1^{*}}(t)}{S_1^{*}(t\mid X)G_1^{*}(t\mid X)}\right\}\right. \notag \\
    &&\quad \ \left. -\frac{A-e^{*}(X)}{e^{*}(X)}\mu_1^{*}(X)S_1^{*}(k\mid X)\right] \label{eqn::dr_to_prove}
\end{eqnarray}
if either $\{e^{*}(X)=e(X), G_1^{*}(t\mid X)=G_1(t\mid X)\}$ or $\{\mu_1^{*}(X)=\mu_1(X), S_1^{*}(t\mid X)=S_1(t\mid X)\}$ for $t\leq k$. Define
\begin{eqnarray*}
    \mathcal{T}_1 &=& \frac{A}{e^{*}(X)}\frac{Y1(T\wedge C>k)}{G_1^*(k\mid X)} -\frac{A-e^{*}(X)}{e^{*}(X)}\mu_1^{*}(X)S_1^{*}(k\mid X) - \mu_1(X)S_1(k\mid X), \\
    \mathcal{T}_2 &=& \frac{A}{e^{*}(X)}\mu_1^{*}(X)S_1^{*}(k\mid X)\int_{0}^{\tilde{T}}\frac{\d M_{G_1^{*}}(t)}{S_1^{*}(t\mid X)G_1^{*}(t\mid X)}.
\end{eqnarray*}
By the identification formula~\eqref{eqn::id_out}, to prove~\eqref{eqn::dr_to_prove}, it suffices to show $E(\mathcal{T}_1+\mathcal{T}_2)=0$ if either $\{e^{*}(X)=e(X), G_1^{*}(t\mid X)=G_1(t\mid X)\}$ or $\{\mu_1^{*}(X)=\mu_1(X), S_1^{*}(t\mid X)=S_1(t\mid X)\}$ for $t\leq k$.

First, we have 
\begin{eqnarray*}
    E\{\d M_{G_1^{*}}(t)\mid X\} &=& E\{1(C\in \d t, \Delta = 0)\mid X\} - E\{1(\tilde{T} \geq t) \d \Lambda_{1}^{*}(t\mid X)\mid X\} \\
    &=& S_1(t\mid X) G_1(t\mid X) \d \Lambda_{1}(t\mid X) - S_1(t\mid X) G_1(t\mid X) \d \Lambda_{1}^{*}(t\mid X) \\
    &=& S_1(t\mid X) G_1(t\mid X) \{\d\Lambda_{1}(t\mid X) - \d\Lambda_{1}^{*}(t\mid X)\},
\end{eqnarray*}
and therefore,
\begin{eqnarray*}
    E(\mathcal{T}_2\mid X) &=& \frac{e(X)}{e^{*}(X)}\mu_1^{*}(X)S_1^{*}(k\mid X)E\left\{\int_{0}^{k}\frac{\d M_{G_1^{*}}(t)}{S_1^{*}(t\mid X)G_1^{*}(t\mid X)} \mid X\right\} \\
    &=& \frac{e(X)}{e^{*}(X)}\mu_1^{*}(X)S_1^{*}(k\mid X) \int_{0}^{k} \frac{S_1(t\mid X) G_1(t\mid X)}{S_1^{*}(t\mid X)G_1^{*}(t\mid X)}\{\d\Lambda_{1}(t\mid X) - \d\Lambda_{1}^{*}(t\mid X)\},
\end{eqnarray*}
where the first equality follows from Assumption~\ref{assump::randomization}. Next, we have
\begin{eqnarray*}
    E(\mathcal{T}_1\mid X) &=& \frac{e(X)G_1(k\mid X)}{e^{*}(X)G_1^{*}(k\mid X)}\mu_1(X)S_1(k\mid X) -\frac{e(X)-e^{*}(X)}{e^{*}(X)}\mu_1^{*}(X)S_1^{*}(k\mid X) - \mu_1(X)S_1(k\mid X) \\
    &=& \frac{e(X)}{e^{*}(X)}\left\{\frac{G_1(k\mid X)}{G_1^{*}(k\mid X)} - 1\right\}\mu_1^{*}(X)S_1^{*}(k\mid X) \\
    &&+ \left\{\frac{e(X)G_1(k\mid X)}{e^{*}(X)G_1^{*}(k\mid X)} - 1\right\}\left\{\mu_1(X)S_1(k\mid X) - \mu_1^{*}(X)S_1^{*}(k\mid X)\right\} \\
    &=& -\frac{e(X)}{e^{*}(X)}\mu_1^{*}(X)S_1^{*}(k\mid X)\int_0^k\frac{G_1(t\mid X)}{G_1^{*}(t\mid X)}\{\d\Lambda_{1}(t\mid X) - \d\Lambda_{1}^{*}(t\mid X)\} \\
    &&+ \left\{\frac{e(X)G_1(k\mid X)}{e^{*}(X)G_1^{*}(k\mid X)} - 1\right\}\left\{\mu_1(X)S_1(k\mid X) - \mu_1^{*}(X)S_1^{*}(k\mid X)\right\},
\end{eqnarray*}
where the last equality follows from the Duhamel equation \citep{gill1990survey, westling2024inference}. Combing $\mathcal{T}_1$ and $\mathcal{T}_2$, we have
\begin{eqnarray*}
    E(\mathcal{T}_1+\mathcal{T}_2) &=& E\{E(\mathcal{T}_1+\mathcal{T}_2\mid X)\} \\
    &=& \frac{e(X)}{e^{*}(X)}\mu_1^{*}(X)S_1^{*}(k\mid X)\int_0^k\left\{\frac{S_1(t\mid X)}{S_1^{*}(t\mid X)} - 1 \right\}\frac{G_1(t\mid X)}{G_1^{*}(t\mid X)}\{\d\Lambda_{1}(t\mid X) - \d\Lambda_{1}^{*}(t\mid X)\} \\
    &&+ \left\{\frac{e(X)G_1(k\mid X)}{e^{*}(X)G_1^{*}(k\mid X)} - 1\right\}\left\{\mu_1(X)S_1(k\mid X) - \mu_1^{*}(X)S_1^{*}(k\mid X)\right\},
\end{eqnarray*}
which is equal to 0 if either $\{e^{*}(X)=e(X), G_1^{*}(t\mid X)=G_1(t\mid X)\}$ or $\{\mu_1^{*}(X)=\mu_1(X), S_1^{*}(t\mid X)=S_1(t\mid X)\}$ for $t\leq k$. 

By symmetry, the part corresponding to the control arm $a=0$ also holds, thus the double robustness in Theorem~\ref{thm::double_robustness} holds.
\QEDB

\subsection{Proof of Theorem~\ref{thm::eif_asym_dist}}
Following the von Mises expansion \citep{hines2022demystifying}, let $P$ and $P_n$ denote the true and empirical distribution of the observed data, respectively, and $\hat P_n$ denotes an estimated $P$, we have
\begin{eqnarray*}
    n^{1/2} (\hat\tau^{\textup{eif}} - \tau) \ =\ n^{-1/2}\sumn D_{\tau}(O_i) + n^{1/2}(P_n - P)\{D_{\tau}(O, \hat P_n) - D_{\tau}(O)\} - n^{1/2}R(P,\hat P_n),
\end{eqnarray*}
where $D_{\tau}(O, \hat P_n)$ is the EIF for $\tau$ when plugging in the estimated values of the nuisance parameters and $R(P,\hat P_n)$ denotes the higher order remainder term. Under Assumption~\ref{assump::donsker}, the empirical process term $n^{1/2}(P_n - P)\{D_{\tau}(O, \hat P_n) - D_{\tau}(O)\}=o_P(1)$ \citep{van2000asymptotic}. The remainder term corresponding to the treatment arm $a=1$ satisfies
\begin{eqnarray*}
    R_1(P,\hat P_n) &=& -E\{D_{\tau}(O, \hat P_n)\} - E\{\hat\mu_1(X)\hat S_1(k\mid X) - \mu_1(X)S_1(k\mid X)\} \\
    &=& -E\left[\frac{A}{\hat e(X)}\left\{\frac{Y1(T\wedge C>k)}{\hat G_1(k\mid X)}+\hat\mu_1(X)\hat S_1(k\mid X)\int_0^k \frac{\d M_{\hat G_1}(t)}{\hat G_1(t\mid X)\hat S_1(t\mid X)}\right\}\right. \\
    &&\quad \quad \  \left. - \frac{A-\hat e(X)}{\hat e(X)}\hat\mu_1(X)\hat S_1(k\mid X) - \hat\mu_1(X)\hat S_1(k\mid X) + \hat\mu_1(X)\hat S_1(k\mid X) - \mu_1(X)S_1(k\mid X)\right] \\
    &=& -E\left[\frac{e(X)}{\hat e(X)}\hat\mu_1(X)\hat S_1(k\mid X)\int_0^k\left\{\frac{S_1(t\mid X)}{\hat S_1(t\mid X)} - 1 \right\}\frac{G_1(t\mid X)}{\hat G_1(t\mid X)}\{\d\Lambda_{1}(t\mid X) - \d\hat\Lambda_{1}(t\mid X)\}\right] \\
    &&+ E\left[\left\{\frac{e(X)G_1(k\mid X)}{\hat e(X)\hat G_1(k\mid X)} - 1\right\}\left\{\mu_1(X)S_1(k\mid X) - \hat\mu_1(X)\hat S_1(k\mid X)\right\}\right] \\
    &=& E\left\{\frac{e(X)}{\hat e(X)}\hat\mu_1(X)\hat S_1(k\mid X)\int_0^k\frac{S_1(t\mid X)-\hat S_1(t\mid X)}{\hat S_1(t\mid X)}\frac{G_1(t\mid X)-\hat G_1(t\mid X)}{\hat G_1(t\mid X)}\d t\right\} \\
    &&+ E\left[\frac{e(X)G_1(k\mid X)-\hat e(X)\hat G_1(k\mid X)}{\hat e(X)\hat G_1(k\mid X)} \left\{\mu_1(X)S_1(k\mid X) - \hat\mu_1(X)\hat S_1(k\mid X)\right\}\right],
\end{eqnarray*}
where the third equality follows from a similar derivation as in the Proof of Theorem~\ref{thm::double_robustness} and the fourth equality follows from the Duhamel equation \citep{gill1990survey, westling2024inference}. Let $R_{11}$ and $R_{12}$ denote the two terms in the last two lines, respectively. For the remainder term to be of small order, we need $R_{11}=o_P(n^{-1/2})$ and $R_{12}=o_P(n^{-1/2})$. 

Next, we show that Assumption~\ref{assump::rate_nuisance} is a sufficient condition for $R_{11}=o_P(n^{-1/2})$ and $R_{12}=o_P(n^{-1/2})$. In the following discussion, we suppress the dependency of the nuisance functions and their estimated values on $X$ for ease of notation. By the facts that $e\hat \mu_1 \hat S_1(k)/\hat e$ is bounded and $1/\{\hat S_1(t)\hat G_1(t)\}$ is bounded for any $t\leq k$, we employ the Cauchy--Schwarz inequality to upper bound the term $R_{11}$ by a constant times $$\int_0^k ||S_1(t)-\hat S_1(t)||_2 ||G_1(t)-\hat G_1(t)||_2 \d t,$$ where $||\cdot||_2$ denotes the $L_2(P)$ norm. It follows that Assumption~\ref{assump::rate_nuisance} guarantees $R_{11}=o_P(n^{-1/2})$. For $R_{12}$, we further suppress the dependency of $S_1(k\mid X)$, $G_1(k\mid X)$ and their estimators on both $k$ and $X$, and rewrite the term as $$R_{12}=E[\{\hat e(G_1-\hat G_1) + G_1(e-\hat e)\}\{\hat\mu_1(S_1-\hat S_1)+S_1(\mu_1-\hat\mu_1)\}/(\hat e \hat G_1)].$$ Because $\hat e/\hat G_1$ is bounded, by the Cauchy-Schwarz inequality, the first term in $R_{12}$ is upper bounded by $||G_1-\hat G_1||_2||S_1-\hat S_1||_2=o_P(n^{-1/2})$ by Assumption~\ref{assump::rate_nuisance}. The other three terms in $R_{12}$ can be similarly bounded. Thus, $R_1(P,\hat P_n)=o_P(n^{-1/2})$. 

Symmetric arguments imply the analogous term for the control arm satisfies $R_0(P,\hat P_n)=o_P(n^{-1/2})$. Therefore, the results in Theorem~\ref{thm::eif_asym_dist} follow from the central limit theorem.
\QEDB

\newpage

\end{appendix}

\end{document}